\documentstyle[art12]{article} 
\author{U.G\"unther\thanks{e-mail: guenther@pool.hrz.htw-zittau.de},
A.Zhuk\thanks{e-mail: zhuk@paco.odessa.ua}\\
Department of Physics, University of Odessa, \\
2 Petra Velikogo St., Odessa 270100, Ukraine}
\title{Stable compactification  and gravitational excitons 
from extra dimensions 
\thanks{Extended version of a lecture given 
at the conference {\it Modern Modified Theories 
of Gravitation and Cosmology},\newline 
Ben Gurion University, Beer Sheva,  June 29-30, 1997.}
}
\font\msbm=msbm10
\def\RR{\hbox{\msbm R}}
\def\NN{\hbox{\msbm N}}

\def\QQ{\hbox{\msbm Q}}

\date{26.09.1997
}
\def\stackunder#1#2{\mathrel{\mathop{#2}\limits_{#1}}}
\catcode`\@=11
\def\Let@{\relax\iffalse{\fi\let\\=\cr\iffalse}\fi}
\def\vspace@{\def\vspace##1{\crcr\noalign{\vskip##1\relax}}}
\def\multilimits@{\bgroup\vspace@\Let@
 \baselineskip\fontdimen10 \scriptfont\tw@
 \advance\baselineskip\fontdimen12 \scriptfont\tw@
 \lineskip\thr@@\fontdimen8 \scriptfont\thr@@
 \lineskiplimit\lineskip
 \vbox\bgroup\ialign\bgroup\hfil$\m@th\scriptstyle{##}$\hfil\crcr}
\def\Sb{_\multilimits@}
\def\endSb{\crcr\egroup\egroup\egroup}
\def\Sp{^\multilimits@}

\newcommand{\be}[1]{\begin{equation}\label{#1}}
\newcommand{\ee}{\end{equation}}
\newcommand{\ba}[1]{\begin{eqnarray}\label{#1}}
\newcommand{\ea}{\end{eqnarray}}

\begin{document}

\maketitle
\renewcommand{\baselinestretch}{0.9} %$\ \ $ \vspace{1cm}\\

\abstract{
We study inhomogeneous multidimensional cosmological models with a higher
dimensional space-time manifold $M = M_0\times\prod\nolimits_{i=1}
^nM_i$  $( n \ge 1 )$  under dimensional reduction
to $D_0$ - dimensional effective models.
Stability due to different types of effective potentials is analyzed
for specific configurations of internal spaces.
Necessary restrictions on the parameters of the models are found and
masses of gravitational excitons
(small inhomogeneous excitations of the scale factors
of the internal spaces near minima of effective potentials) are calculated.
}

\bigskip

\hspace*{0.950cm} PACS number(s): 04.50.+h, 98.80.Hw 
%%%%%%%%%%%%%%%%%%%%%%%%%%%%%%%%%%%%%%%%%%%%%%%%%%%%%%%%%%%%%%%%%
%
\newpage

\section{Introduction}

\setcounter{equation}{0}

Many modern theories beyond the Standard Model include the hypothesis that
our space-time has a dimensionality of more than four. String theory \cite
{1a} and its recent generalizations --- p-brane, M- and F-theory \cite
{1aa,1b,1c} widely use this concept and give it a new foundation. The most
consistent formulations of these theories are possible in space-times with
critical dimensions $D_c>4$, for example, in string theory there are $D_c=26$
or $10$ for the bosonic and supersymmetric version, respectively. Usually it
is supposed that a $D$-dimensional manifold $M$ undergoes a ''spontaneous
compactification'' \cite{c1}-\cite{c4}: 
\mbox{$M\to M^4\times B^{D-4}$,} where $M^4
$ is the 4-dimensional external space-time and $B^{D-4}$ is a compact
internal space. It is clear that such compactifications necessarily lead to
cosmological consequences. One way to study them is to investigate
simplified multidimensional cosmological models (MCM) with topology 
\begin{equation}
\label{1.1}M=M_0\times M_1\times \dots \times M_n,
\end{equation}
where $M_0$ denotes the $D_0$ - dimensional (usually $D_0=4$) external
space-time and $M_i$ \quad $(i=1,\dots ,n)$ are $D_i$ - dimensional internal
spaces. To make the internal dimensions unobservable at present time the
internal spaces have to be compact and reduced to scales near Planck length $%
L_{Pl}\sim 10^{-33}cm$, i.e. scale factors $a_i$ of the internal spaces
should be of order of $L_{Pl}$. In this case we cannot move in
extra-dimensions and our space-time is apparently 4-dimensional. 

Compact internal spaces can exist for any sign of scalar curvature \cite{2a}.
There is no problem to construct compact spaces with  positive curvature 
\cite{p1,p2}. (For example, every Einstein manifold with constant positive
curvature is necessarily compact \cite{p3}.) However, Ricci-flat spaces and
negative curvature spaces can also be compactified. This can be achieved by
appropriate periodicity conditions for the coordinates \cite{2}-\cite{5a}
or, equivalently, through the action of discrete groups $\Gamma $ of
isometries related to face pairings and to the manifold's topology. For
example, 3-dimensional spaces of constant negative curvature are isometric
to the open, simply connected, infinite hyperbolic (Lobachevsky) space $H^3$ 
\cite{p1,p2}. But there exist also an infinite number of compact, multiply
connected, hyperbolic coset manifolds 
\mbox{$H^3/\Gamma $,} which can be used for
the construction of FRW metrics with negative curvature \cite{2,4}. These
manifolds are built from a fundamental polyhedron (FP) in $H^3$ with faces
pairwise identified. The FP determines a tessellation of $H^3$ into cells
which are replicas of the FP, through the action of the discrete group $%
\Gamma $ of isometries \cite{4}. In \cite{2a} it is, e.g., shown that by
this way one can construct tori with genus two or more. The simplest example
of Ricci-flat compact spaces is given by $D$ - dimensional tori $T^D=\RR %
^D/\Gamma $. Thus, internal spaces may have nontrivial global topology,
being compact (i.e. closed and bounded) for any sign of spatial curvature.

In the cosmological context, internal spaces can be called compactified,
when they are obtained by a compactification \cite{k0} or factorization
(''wrapping'') in the usual mathematical understanding (e.g. by replacements
of the type $\RR ^D\rightarrow S^D$, $\RR ^D\rightarrow \RR ^D/\Gamma $ or $%
H^D\rightarrow H^D/\Gamma $) with additional contraction of the sizes to
Planck scale. The physical constants that appear in the effective
4-dimensional theory after dimensional reduction of an originally
higher-dimensional model are the result of integration over the extra
dimensions. If the volumes of the internal spaces would change, so would the
observed constants. Because of limitation on the variability of these
constants \cite{k1,k2} the internal spaces are static or at least slowly
variable since the time of primordial nucleosynthesis and as we mentioned
above their sizes are of the order of the Planck length. Obviously, such
compactifications have to be stable against small fluctuations of the sizes
(the scale factors $a_i$) of the internal spaces. This means that the
effective potential of the model obtained under dimensional reduction to a
4-dimensional effective theory should have minima at $a_i\sim L_{Pl}$ $%
(i=1,\dots ,n)$. Because of its crucial role the problem of stable
compactification of extra dimensions was intensively studied in a large
number of papers \cite{6}-\cite{22}. As result certain conditions were
obtained which ensure the stability of these compactifications. However,
position of a system at a minimum of an effective potential means not
necessarily that extra-dimensions are unobservable. As we shall show below,
small excitations of a system near a minimum can be observed as massive
scalar fields in the external space-time. In solid state physics,
excitations of electron subsystems in crystals are called excitons. In our
case the internal spaces are an analog of the electronic subsystem and their
excitations can be called gravitational excitons. If masses of these
excitations are much less than Planck mass $M_{Pl}\sim 10^{-5}g$, they
should be observable confirming the existence of extra-dimensions. In the
opposite case of very heavy excitons with masses $m\sim M_{Pl}$ it is
impossible to excite them at present time and extra-dimensions are
unobservable by this way.

%%%%%%%%%%%%%%%%%%%%%%%%%%%%%%%%%%%%%%%%%%%%%%%%%%%%%%%%%%%%%%%%%
%
\newpage

\section{The model}

\setcounter{equation}{0}

We consider a cosmological model with metric 
\begin{equation}
\label{2.1}g=g^{(0)}+\sum_{i=1}^ne^{2\beta ^i(x)}g^{(i)}, 
\end{equation}
which is defined on manifold (\ref{1.1}) where $x$ are some coordinates of
the $D_0$ - dimensional manifold $M_0$ and 
\begin{equation}
\label{2.2}g^{(0)}=g_{\mu \nu }^{(0)}(x)dx^\mu \otimes dx^\nu . 
\end{equation}
Let manifolds $M_i$ be $D_i$ - dimensional Einstein spaces with metric $%
g^{(i)},$  i.e. 
\begin{equation}
\label{2.3}R_{mn}\left[ g^{(i)}\right] =\lambda ^ig_{mn}^{(i)},\qquad
m,n=1,\ldots ,D_i 
\end{equation}
and 
\begin{equation}
\label{2.4}R\left[ g^{(i)}\right] =\lambda ^iD_i\equiv R_i. 
\end{equation}
In the case of constant curvature spaces parameters $\lambda ^i$ are
normalized as $\lambda ^i=k_i(D_i-1)$ with $k_i=\pm 1,0$. We note that each
of the spaces $M_i$ can be split into a product of Einstein spaces: $%
M_i\rightarrow \prod_{k=1}^{n_i}M_i^k$ \cite{23}. Here $M_i^k$ are Einstein
spaces of dimensions $D_i^k$ with metric $g_{(k)}^{(i)}$: $R_{mn}\left[
g_{(k)}^{(i)}\right] =\lambda _k^ig_{(k)mn}^{(i)}\quad (m,n=1,\ldots ,D_i^k)$
and $R\left[ g_{(k)}^{(i)}\right] =\lambda _k^iD_i^k$. Such a splitting
procedure is well defined provided $M_i^k$ are not Ricci - flat \cite{23,24}%
. If $M_i$ is a split space, then for curvature and dimension we have
respectively \cite{23}: $R\left[ g^{(i)}\right] =\sum_{k=1}^{n_i}R\left[
g_{(k)}^{(i)}\right] $ and $D_i=\sum_{k=1}^{n_i}D_i^k$. Later on we shall
not specify the structure of the spaces $M_i$. We require only $M_i$ to be
compact spaces with arbitrary sign of curvature.

With total dimension $D=\sum_{i=0}^nD_i$, $\kappa ^2$ a $D$ - dimensional
gravitational constant, $\Lambda $ - a $D$ - dimensional cosmological
constant and $S_{YGH}$ the standard York - Gibbons - Hawking boundary term 
\cite{24a,24b}, we consider an action of the form 
\begin{equation}
\label{2.5}S=\frac 1{2\kappa ^2}\int\limits_Md^Dx\sqrt{|g|}\left\{
R[g]-2\Lambda \right\} +S_{add}+S_{YGH}. 
\end{equation}
The additional potential term 
\begin{equation}
\label{2.6}S_{add}=-\int\limits_Md^Dx\sqrt{|g|}\rho (x) 
\end{equation}
is not specified and left in its general form, taking into account the
Casimir effect \cite{6}, the Freund - Rubin monopole ansatz \cite{c2}, a
perfect fluid \cite{26,27} or other hypothetical potentials \cite{20,22}. In
all these cases $\rho $ depends on the external coordinates through the
scale factors $a_i(x)=e^{\beta ^i(x)}\ (i=1,\ldots ,n)$ of the internal
spaces. We did not include into the action (\ref{2.5}) a minimally coupled
scalar field with potential $U(\psi )$, because in this case there exist no
solutions with static internal spaces for scalar fields $\psi $ depending on
the external coordinates \cite{20}.

After dimensional reduction the action reads 
$$
S=\frac 1{2\kappa _0^2}\int\limits_{M_0}d^{D_0}x\sqrt{|g^{(0)}|}%
\prod_{i=1}^ne^{D_i\beta ^i}\left\{ R\left[ g^{(0)}\right] -G_{ij}g^{(0)\mu
\nu }\partial _\mu \beta ^i\,\partial _\nu \beta ^j+\right. 
$$
\begin{equation}
\label{2.7}+\sum_{i=1}^n\left. R\left[ g^{(i)}\right] e^{-2\beta
^i}-2\Lambda -2\kappa ^2\rho \right\} , 
\end{equation}
where $\kappa _0^2=\kappa ^2/\mu $ is the $D_0$ - dimensional gravitational
constant,\\ $\mu =\prod_{i=1}^n\mu _i=\prod_{i=1}^n\int\limits_{M_i}d^{D_i}y 
\sqrt{|g^{(i)}|}$ and $G_{ij}=D_i\delta _{ij}-D_iD_j$ \\ $(i,j=1,\ldots ,n)$ is
the midisuperspace metric \cite{28,29}. Here the scale factors $\beta ^i$ of
the internal spaces play the role of scalar fields. Comparing this action
with the tree-level effective action for a bosonic string it can be easily
seen that the volume of the internal spaces $e^{-2\Phi }\equiv
\prod_{i=1}^ne^{D_i\beta ^i}$ plays the role of the dilaton field \cite
{23,29,30}. We note that sometimes all scalar fields associated with $\beta
^i$ are called dilatons. Action (\ref{2.7}) is written in the Brans - Dicke
frame. Conformal transformation to the Einstein frame 
\begin{equation}
\label{2.8}\hat g_{\mu \nu }^{(0)}=e^{-\frac{4\Phi }{D_0-2}}g_{\mu \nu
}^{(0)}={\left( \prod_{i=1}^ne^{D_i\beta ^i}\right) }^{\frac 2{D_0-2}}g_{\mu
\nu }^{(0)} 
\end{equation}
yields 
\begin{equation}
\label{2.9}S=\frac 1{2\kappa _0^2}\int\limits_{M_0}d^{D_0}x\sqrt{|\hat
g^{(0)}|}\left\{ \hat R\left[ \hat g^{(0)}\right] -\bar G_{ij}\hat g^{(0)\mu
\nu }\partial _\mu \beta ^i\,\partial _\nu \beta ^j-2U_{eff}\right\} . 
\end{equation}
The tensor components of the midisuperspace metric (target space metric on $%
\RR _T^n$ ) $\bar G_{ij}\ (i,j=1,\ldots ,n)$, its inverse metric $\bar
G^{ij}$ and the effective potential are respectively 
\begin{equation}
\label{2.10}\bar G_{ij}=D_i\delta _{ij}+\frac 1{D_0-2}D_iD_j, 
\end{equation}
\begin{equation}
\label{2.11}\bar G^{ij}=\frac{\delta ^{ij}}{D_i}+\frac 1{2-D} 
\end{equation}
and 
\begin{equation}
\label{2.12}U_{eff}={\left( \prod_{i=1}^ne^{D_i\beta ^i}\right) }^{-\frac
2{D_0-2}}\left[ -\frac 12\sum_{i=1}^nR_ie^{-2\beta ^i}+\Lambda +\kappa
^2\rho \right] . 
\end{equation}
We remind that $\rho $ depends on the scale factors of the internal spaces: $%
\rho =\rho \left( \beta ^1,\ldots ,\beta ^n\right) $. Thus, we are led to
the action of a self-gravitating $\sigma -$model with flat target space $(%
\RR _T^n,\bar G)$ (\ref{2.10}) and self-interaction described by the
potential (\ref{2.12}).

Let us first consider the case of one internal space: $n=1$. Redefining the
dilaton field as 
\begin{equation}
\label{2.13}\varphi \equiv \pm \sqrt{\frac{D_1(D-2)}{D_0-2}}\beta ^1 
\end{equation}
we get for action and effective potential respectively 
\begin{equation}
\label{2.14}S=\frac 1{2\kappa _0^2}\int d^{D_0}x\sqrt{|\hat g^{(0)}|}\left\{
\hat R\left[ \hat g^{(0)}\right] -\hat g^{(0)\mu \nu }\partial _\mu \varphi
\,\partial _\nu \varphi -2U_{eff}\right\} 
\end{equation}
and 
\begin{equation}
\label{2.15}U_{eff}=e^{2\varphi {\left[ \frac{D_1}{(D-2)(D_0-2)}\right] }%
^{1/2}}\left[ -\frac 12R_1e^{2\varphi {\left[ \frac{D_0-2}{D_1(D-2)}\right] }%
^{1/2}}+\Lambda +\kappa ^2\rho (\varphi )\right] , 
\end{equation}
where in the latter expression we use for definiteness sign minus.

Coming back to the general case $n>1$ we bring midisuperspace metric (target
space metric) (\ref{2.10}) by a regular coordinate transformation 
\begin{equation}
\label{2.16}\varphi =Q\beta ,\quad \beta =Q^{-1}\varphi \ 
\end{equation}
to a pure Euclidean form 
\begin{equation}
\label{2.17} 
\begin{array}{c}
\bar G_{ij}d\beta ^i\otimes d\beta ^j=\sigma _{ij}d\varphi ^i\otimes
d\varphi ^j=\sum_{i=1}^nd\varphi ^i\otimes d\varphi ^i, \\  
\\ 
\bar G=Q^{\prime }Q,\quad \sigma ={\rm diag\ }(+1+1,\ldots ,+1). 
\end{array}
\end{equation}
(The prime denotes the transposition.) An appropriate transformation $Q:\
\beta ^i\mapsto \varphi ^j=Q_i^j\beta ^i$ is given e.g. by \cite{28} 
\begin{equation}
\label{2.18} 
\begin{array}{ll}
\varphi ^1 & =-A\sum_{i=1}^nD_i\beta ^i \\  
&  \\ 
\varphi ^i & =\left[ D_{i-1}/\Sigma _{i-1}\Sigma _i\right]
^{1/2}\sum_{j=i}^nD_j(\beta ^j-\beta ^{i-1}) 
\end{array}
\end{equation}
where $i=2,\ldots ,n$, $\Sigma _i=\sum_{j=i}^nD_j$, 
\begin{equation}
\label{2.20}A=\pm {\left[ \frac 1{D^{\prime }}\frac{D-2}{D_0-2}\right] }%
^{1/2}, 
\end{equation}
and $D^{\prime }=\sum_{i=1}^nD_i$. So we can write action (\ref{2.9}) as 
\begin{equation}
\label{2.21}S=\frac 1{2\kappa _0^2}\int\limits_{M_0}d^{D_0}x\sqrt{|\hat
g^{(0)}|}\left\{ \hat R\left[ \hat g^{(0)}\right] -\sigma _{ik}\hat
g^{(0)\mu \nu }\partial _\mu \varphi ^i\,\partial _\nu \varphi
^k-2U_{eff}\right\} 
\end{equation}
with effective potential 
\begin{equation}
\label{2.22}U_{eff}=e^{\frac 2{A(D_0-2)}\varphi ^1}\left( -\frac
12\sum_{i=1}^nR_ie^{-2{(Q^{-1})^i}_k\varphi ^k}+\Lambda +\kappa ^2\rho
\right) . 
\end{equation}

%%%%%%%%%%%%%%%%%%%%%%%%%%%%%%%%%%%%%%%%%%%%%%%%%%%%%%%%%%%%%%%%%
%

\section{Gravitational excitons as solutions of a linear $\sigma $-model%
\label{mark1}}

\setcounter{equation}{0}

In general, the effective potential (\ref{2.22}) is a highly nonlinear
function and it would be rather difficult to obtain explicit solutions $%
\varphi ^i$ of the Euler-Lagrange-equation for the corresponding $\sigma $%
-model action (\ref{2.21}) analytically. The situation crucially simplifies,
when we are interested in small field fluctuations $\xi ^i$ around the
minima of potential (\ref{2.22}) only.

Let us suppose that these minima are localized at points $\vec \varphi
_c,c=1,...,m$%
\begin{equation}
\label{3.1} 
\begin{array}{l}
\left. 
\frac{\partial U_{eff}}{\partial \varphi ^i}\right| _{\vec \varphi _c}=0\
,\quad \ \xi ^i\equiv \varphi ^i-\varphi _{(c)}^i, \\  \\ 
U_{eff}=U_{eff}\left( \vec \varphi _c\right) +\frac 12\sum_{i,k=1}^n\bar
a_{(c)ik}\xi ^i\xi ^k+O(\xi ^i\xi ^k\xi ^l) 
\end{array}
\end{equation}
and that the Hessians 
\begin{equation}
\label{3.2}\bar a_{(c)ik}:=\left. \frac{\partial ^2U_{eff}}{\partial \xi
^i\,\partial \xi ^k}\right| _{\vec \varphi _c} 
\end{equation}
are not vanishing identically. The action functional (\ref{2.21}) reduces
then to a family of action functionals for the fluctuation fields $\xi ^i$%
\begin{equation}
\label{3.3} 
\begin{array}{ll}
S_{(c)}= & \frac 1{2\kappa _0^2}\int\limits_{M_0}d^{D_0}x 
\sqrt{|\hat g^{(0)}|}\left\{ \hat R\left[ \hat g^{(0)}\right]
-2U_{eff}\left( \vec \varphi _c\right) -\right. \\  &  \\  
& \left. -\sigma _{ik}\hat g^{(0)\mu \nu }\partial _\mu \xi ^i\,\partial
_\nu \xi ^k-\bar a_{(c)ik}\xi ^i\xi ^k\right\} ,\ c=1,...,m. 
\end{array}
\end{equation}
It remains to diagonalize the Hessians $\bar a_{(c)ik}$ by appropriate \\ $%
SO(n)-$ro\-ta\-tions $S_c:\ \xi \mapsto \psi =S_c\xi ,\quad S_c^{\prime
}=S_c^{-1}$%
\begin{equation}
\label{3.4}\bar A_c=S_c^{\prime }M_c^2S_c,\quad M_c^2={\rm diag\ }%
(m_{(c)1}^2,m_{(c)2}^2,\ldots ,m_{(c)n}^2), 
\end{equation}
leaving the kinetic term $\sigma _{ik}\hat g^{(0)\mu \nu }\partial _\mu \xi
^i\,\partial _\nu \xi ^k$ invariant 
\begin{equation}
\label{3.5}\sigma _{ik}\hat g^{(0)\mu \nu }\partial _\mu \xi ^i\,\partial
_\nu \xi ^k=\sigma _{ik}\hat g^{(0)\mu \nu }\partial _\mu \psi ^i\,\partial
_\nu \psi ^k, 
\end{equation}
and we arrive at action functionals for decoupled normal modes of li-\\ near $%
\sigma -$models in the background metric $\hat g^{(0)}$ of the external
\mbox{space-time:} 
\begin{eqnarray}\label{3.6}
S & = & \frac{1}{2\kappa _0^2}\int \limits_{M_0}d^{D_0}x \sqrt
{|\hat g^{(0)}|}\left\{\hat R\left[\hat g^{(0)}\right] - 2\Lambda
_{(c)eff}\right\} + \nonumber\\
\ & + & \sum_{i=1}^{n}\frac{1}{2}\int \limits_{M_0}d^{D_0}x \sqrt
{|\hat g^{(0)}|}\left\{-\hat g^{(0)\mu \nu}\psi ^i_{,\mu}\psi
^i_{,\nu} -
m_{(c)i}^2\psi ^i\psi ^i\right\}\ , 
\end{eqnarray}where $c=1,...,m,\quad \Lambda _{(c)eff}\equiv U_{eff}\left(
\vec \varphi _c\right) $ and the factor $\sqrt{\mu /\kappa ^2}$ has been
included into $\psi $ for convenience: $\sqrt{\mu /\kappa ^2}\psi
\rightarrow \psi $.

Thus, conformal excitations of the metric of the internal spaces behave as
massive scalar fields developing on the background of the external space -
time. By analogy with excitons in solid state physics where they are
excitations of the electronic subsystem of a crystal, the excitations of the
internal spaces may be called gravitational excitons.

Before we turn to a discussion of concrete classes of effective potentials
and physical conditions on the parameters of the model, which must be
fulfilled for compatibility with observational data from our present time
Universe, we consider some general features of the model.

First we note, that according to expansion (\ref{3.1}) for $\bar A_c\neq 0$
and up to second order in $\xi ^i$, the effective potential (\ref{2.22})
has a minimum at a point $\vec \varphi _c$ iff 
\begin{equation}
\label{3.7}\xi ^{\prime }\bar A_c\xi \equiv \sum_{i,k=1}^n\bar a_{(c)ik}\xi
^i\xi ^k\geq 0,\quad \forall \xi ^k, 
\end{equation}
with exception of $\xi $$^1=\xi ^2=\ldots =\xi ^n=0$. This condition is
equivalent to the requirement that at least one of the exciton masses should
be strictly positive, whereas the remaining could vanish 
\begin{equation}
\label{3.8}m_{(c)i}^2\geq 0,\ m_{(c)k}^2>0\ {\rm for\ at\ least\ one\ }k. 
\end{equation}
In the following sections we focus on models with strictly positive exciton
masses $m_{(c)i}^2>0,\ \forall i.$ In this case, according to the
Sylvester criterion, positivity of the quadratic form (\ref{3.7}) is assured
by the positivity of the principal minors of the matrix $\bar A_c:$%
\begin{equation}
\label{3.9} 
\begin{array}{l}
\bar a_{(c)11}>0,\quad \left| 
\begin{array}{cc}
\bar a_{(c)11} & \bar a_{(c)12} \\ 
\bar a_{(c)21} & \bar a_{(c)22} 
\end{array}
\right| >0,\quad \ldots \\  
\\ 
\ldots ,\quad \left| 
\begin{array}{ccc}
\bar a_{(c)11} & \cdots & \bar a_{(c)1n} \\ 
\bar a_{(c)21} & \cdots & \bar a_{(c)2n} \\ 
\cdots & \cdots & \cdots \\ 
\bar a_{(c)n1} & \cdots & \bar a_{(c)nn} 
\end{array}
\right| =\det \bar A_c>0. 
\end{array}
\end{equation}
The consideration of Mexican-hat-type potentials, which correspond to
degenerated minima $(m_{(c)1}^2=0,\ m_{(c)2}^2>0,\ \ldots )$, yielding
massless modes similar to Goldstone bosons we leave for a separate paper.

From a technical point of view, the explicit calculation of the exciton
masses can be considerably simplified if one makes use of the equivalence of 
$\varphi -$representation and $\beta -$representation: Minima in $\varphi -$%
representation correspond to minima in $\beta -$representation. This
property of the model is easily shown: Under the regular linear
transformation (\ref{2.16}) $\varphi =Q\beta $, which depends, according to
(\ref{2.18}) and (\ref{2.20}), on the dimensional structure of the total
midi-superspace $M$ only, extremum condition, Hessian and quadratic form
transform as follows: 
\begin{equation}
\label{3.10}\left. \frac{\partial U_{eff}}{\partial \varphi ^i}\right|
_{\vec \varphi _c}=\left. \frac{\partial U_{eff}}{\partial \beta ^k}\right|
_{\vec \beta _c}{(Q^{-1})^k}_i=0,\quad \varphi _c=Q\beta _c, 
\end{equation}
\begin{equation}
\label{3.11}a_{(c)ik}=\left. \frac{\partial ^2U_{eff}}{\partial \beta
^i\,\partial \beta ^k}\right| _{\vec \beta _c}=\frac{\partial \varphi ^j}{%
\partial \beta ^i}\left. \frac{\partial ^2U_{eff}}{\partial \varphi
^j\,\partial \varphi ^l}\right| _{\vec \varphi _c}\frac{\partial \varphi ^l}{%
\partial \beta ^k}\equiv Q_i^j\bar a_{(c)jl}Q_k^l, 
\end{equation}
\begin{equation}
\label{3.12}\xi =Q\eta ,\quad \eta =Q^{-1}\xi ,\quad \xi ^i\equiv
\varphi ^i-\varphi _c^i,\quad \eta ^i\equiv \beta ^i-\beta _c^i 
\end{equation}
\begin{equation}
\label{3.13}{\eta }^{\prime }A_c\eta =(Q^{-1}\xi )^{\prime }Q^{\prime }\bar
A_cQ(Q^{-1}\xi )=\xi ^{\prime }\bar A_c\xi . 
\end{equation}
This means, first, that extrema in $\varphi -$representation correspond to
extrema in $\beta -$representation. Second, (\ref{3.11}) shows that $A_c$
and $\bar A_c$ are congruent matrices \cite{31}. Hence, their rank and
signature coincide \cite{31}, and positive eigenvalues of $A_c$ correspond
to positive eigenvalues of $\bar A_c$. The equivalence of the
representations is established.

Furthermore, it is easy to see from (\ref{2.17}), (\ref{3.4}) and (\ref{3.11}%
) that eigenvalues of matrices $\bar A_c$ coincide with eigenvalues of
matrices $\bar G^{-1}A_c,$  so that exciton masses can be calculated without
technical problems from the Hessian in $\beta -$representation directly. For
two-scale-factor models $(n=2)$ we have, for example, 
\begin{equation}
\label{3.14}m_{(c)1,2}^2=\frac 12\left[ Tr(B_c)\pm \sqrt{Tr^2(B_c)-4\det
(B_c)}\right] , 
\end{equation}
where 
\begin{equation}
\label{3.15}B_c=\bar A_c\quad {\rm or}\quad B_c=\bar G^{-1}A_c. 
\end{equation}
It can be easily seen that $m_{(c)1}^2,\ m_{(c)2}^2$ are positive iff $%
\bar a_{(c)11},\bar a_{(c)22}>0$ and $\bar a_{(c)11}\bar a_{(c)22}>\bar
a_{(c)12}^2$.

As we will show explicitly in the next sections, models of the same type,
e.g. for a one-component perfect fluid, may be unstable in the case of two
independently varying scale factors $(\beta ^1,\beta ^2)$, but become stable
under scale factor reduction, i.e. when the scale factors are connected by a
constraint $\beta ^1=\beta ^2=\beta $. The reason for this interesting
behavior originates in the form of the effective potential $U_{eff}$ at the
extremum point.

Let us illustrate this situation with a reduction of an $n-$scale-factor
model to a one-scale-factor model. In order to simplify our calculation we
introduce the projection operator $P$ on the one-dimensional constraint
subspace $\RR _P^1=\left\{ \bar \beta =(\beta ^1,\ldots ,\beta ^n)\mid \beta
^1=\beta ^2=\ldots =\beta ^n=\beta \right\} $ of the $n-$dimensional target
space $\RR _T^n$ of the $\sigma -$model: 
\begin{equation}
\label{3.11a}P\RR _T^n=\RR _P^1\subset \RR _T^n. 
\end{equation}
Explicitly this projection operator can be constructed from the normalized
base vector $\bar e$ of the subspace \mbox{$\RR _P^1.$} With 
\begin{equation}
\label{3.12a}\bar e=\frac 1{\sqrt{n}}\left( 
\begin{array}{c}
1 \\ 
\vdots \\ 
1 
\end{array}
\right) 
\end{equation}
we have 
\begin{equation}
\label{3.13a}P=\bar e\otimes \bar e^{^{\prime }}=\frac 1n\left( 
\begin{array}{c}
1 \\ 
\vdots \\ 
1 
\end{array}
\right) \otimes \left( 
\begin{array}{ccc}
1 & \cdots & 1 
\end{array}
\right) =\frac 1n\left( 
\begin{array}{ccc}
1 & \cdots & 1 \\ 
\vdots &  & \vdots \\ 
1 & \cdots & 1 
\end{array}
\right) 
\end{equation}
and $P^2=P$.

Let us now calculate the exciton mass $m_{(c)0}$ for the reduced model. For
this purpose we introduce the exciton Lagrangian, written according to (\ref
{2.9}), (\ref{2.17}), (\ref{3.6}) and (\ref{3.13}) in terms of the
fluctuation fields $\bar \eta =(\eta ^1,\ldots ,\eta ^n),\ \eta ^i\equiv
\beta ^i-\beta _c^i$ 
\begin{equation}
\label{3.14a}{\cal L}_{exci}=-\left[ \bar \eta \bar G\widehat{K}\bar \eta
+\bar \eta A_{(c)}\bar \eta \right] . 
\end{equation}
$\widehat{K}:=\overleftarrow{\partial }_\mu \widehat{g}^{(o)\mu \nu } 
\overrightarrow{\partial }_\nu $ denotes the pure kinetic operator. Under
scale factor reduction $\bar \eta =(\eta ,\ldots ,\eta )$ this Lagrangian
simplifies to 
\begin{equation}
\label{3.15a}{\cal L}_{exci}=-\left[ \gamma _1\eta \widehat{K}\eta +\gamma
_{(c)2}\eta ^2\right] ,\quad 
\end{equation}
\begin{equation}
\label{3.16a}\gamma _1:=n\bar e^{^{\prime }}\bar G\bar e=\sum_{i,j}\bar
G_{ij},\quad \gamma _{(c)2}:=n\bar e^{^{\prime }}A_{(c)}\bar
e=\sum_{i,j}A_{(c)ij} 
\end{equation}
so that the substitution $\eta =\gamma _1^{-1/2}\psi $ yields the effective
one-scale-factor Lagrangian ${\cal L}_{exci}=-\left[ \psi \widehat{K}\psi
+\psi m_{(c)0}^2\psi \right] $ with exciton mass \\ $m_{(c)0}^2=\gamma
_{(c)2}/\gamma _1.$  Taking into account that $\bar e^{^{\prime
}}A_{(c)}\bar e=Tr\left[ PA_{(c)}\right] ,$  \ $A_{(c)}=Q^{^{\prime
}}S_c^{^{\prime }}M_{(c)}^2S_cQ$ and $M_{(c)}^2=diag(m_{(c)1}^2,\ldots
,m_{(c)n}^2)$ the needed relation between the exciton masses of the reduced
and unreduced $n-$scale-factor models is now easily established as 
\begin{equation}
\label{3.17a}m_{(c)0}^2=n\gamma _1^{-1}Tr\left[ QPQ^{^{\prime
}}S_c^{^{\prime }}M_{(c)}^2S_c\right] . 
\end{equation}
With use of explicit expressions for transformation matrix $Q$ (\ref{2.18})
and target space metric $\bar G_{ij}$ (\ref{2.10}) we have 
\begin{equation}
\label{3.18a} 
\begin{array}{l}
\left[ Q\bar e\right] _i=- 
\frac{AD^{^{\prime }}}{\sqrt{n}}\delta _{i1},\quad QPQ^{^{\prime }}= 
\frac{D^{^{\prime }}}n\frac{D-2}{D_o-2}\widetilde{P}, \\  \\ 
\widetilde{P}_{ik}:=\delta _{i1}\delta _{k1}\quad {\rm and}\quad \gamma
_1=D^{^{\prime }}\frac{D-2}{D_o-2}, 
\end{array}
\end{equation}
so that relation (\ref{3.17a}) can be finally rewritten in terms of masses
and components of the $SO(n)-$matrices $S_c$ only 
\begin{equation}
\label{3.19}m_{(c)0}^2=Tr\left[ \widetilde{P}S_c^{^{\prime
}}M_{(c)}^2S_c\right] =\sum_{i=1}^n\left( S_{(c)i1}\right) ^2m_{(c)i}^2. 
\end{equation}
In its compact form this mass formula implicitly reflects the behavior of
the effective potential $U_{eff}$ in the vicinity $\Omega _{\vec \beta
_c}\subset \RR _T^n$ of the extremum point $\vec \beta _c$. So, the exciton
masses squared $m_{(c)1}^2,\ldots ,\ m_{(c)n}^2$ describe the potential as
function over the $n-$dimensional $\vec \beta _c-$vicinity $\Omega _{\vec
\beta _c}$, whereas $m_{(c)0}^2$ characterizes $U_{eff}$ as function over
the line interval $\Omega _{\vec \beta _c}\cap \RR _P^1$ only. From (\ref
{3.19}) it is obvious that a positive exciton mass in the reduced model,
corresponding to a minimum of the effective potential over the line interval 
$\Omega _{\vec \beta _c}\cap \RR _P^1$, is not only possible for stable
configurations of the unreduced model $m_{(c)1}^2>0,\ldots ,m_{(c)n}^2>0\ $
, but even in cases when the potential $U_{eff}$ has a saddle point at $\vec
\beta _c$ and the unreduced model is unstable. For the masses we have in
these cases \mbox{$m_{(c)i}^2>0,$} \ $m_{(c)k}^2<0$, for some $i$ and $k\ $, and
massive excitons in the reduced model correspond to exciton - tachyon
configurations in the unreduced model.

As conclusion of this section we want to make a few remarks concerning the
form of the effective potential. From the physical point of view it is clear
that the effective potential should satisfy following conditions:%
\begin{eqnarray}\label{3.19a}
(i)\, \qquad a_{(c)i} & = & e^{\beta_c^i}\,  \mbox{ \small
$^{>}_{\sim}$ }\,L_{Pl},
\nonumber\\
(ii)\ \ \, \quad m_{(c)i} & \leq & M_{Pl},\nonumber\\
(iii) \,  \quad  \Lambda _{(c)eff} & \rightarrow & 0.
\end{eqnarray}
The first condition expresses the fact that the internal spaces should be
unobservable at the present time and stable against quantum gravitational
fluctuations. This condition ensures the applicability of the classical
gravitational equations near positions of minima of the effective potential.
The second condition means that the curvature of the effective potential
should be less than Planckian one. Of course, gravitational excitons can be
excited at the present time if $m_i\ll M_{Pl}$. The third condition reflects
the fact that the cosmological constant at the present time is very small: 
\begin{equation}
\label{3.20}\left| \Lambda \right| \leq 10^{-54}{\mbox{cm}}^{-2}\approx
10^{-120}\Lambda _{Pl}, 
\end{equation}
where $\Lambda _{Pl}=L_{Pl}^{-2}$. Thus, for simplicity, we can demand $%
\Lambda _{eff}=U_{eff}(\vec \beta _c)=0$. (We used the abbreviation $\Lambda
_{eff}\equiv \Lambda _{(c)eff}$.) Strictly speaking, in the multi-minimum
case $(c>1)$ we can demand \mbox{$a_{(c)i}\sim L_{Pl}$} 
and $\Lambda _{(c)eff}=0$
only for one of the minima to which corresponds the present universe state.
For all other minima it may be $a_{(c)i}\gg L_{Pl}$ and $|\Lambda
_{(c)eff}|\gg 0$.

In the following sections we test several types of internal space
configurations and effective potentials on their compatibility with physical
conditions (\ref{3.19a}).

\section{Pure geometrical potentials: $\rho \equiv 0$\label{mark2}}

\setcounter{equation}{0}

In the case of an effective potential of pure geometric type $(\rho \equiv
0) $ the condition for the existence of an extremum $\frac{\partial
U_{eff,0} }{\partial \beta ^k}=0$ implies a fine-tuning 
\begin{equation}
\label{4.1} 
\begin{array}{l}
\frac{R_k}{D_k}e^{-2\beta _c^k}=\frac{2\Lambda }{D-2}\equiv \tilde C,\quad
k=1,\ldots ,n \\  \\ 
\Longrightarrow \qquad e^{\beta _c^k}=\left[ \frac{R_kD_i}{R_iD_k}\right]
^{1/2}e^{\beta _c^i} 
\end{array}
\end{equation}
of the scale factors and $\mbox{\rm sign}\,\Lambda =\mbox{\rm
sign}\,R_i$. With the help of the explicit formula for the target space
metric (\ref{2.10}) we get for the Hessian 
\begin{equation}
\label{4.2} 
\begin{array}{ll}
a_{(c)ik}\equiv \left. \frac{\partial ^2U_{eff,0}}{\partial \beta
^i\,\partial \beta ^k}\right| _{\vec \beta _c} & =- 
\frac{4\Lambda _{eff}}{D_0-2}\left[ \frac{D_iD_k}{D_0-2}+\delta
_{ik}D_k\right] =-\frac{4\Lambda _{eff}}{D_0-2}\bar G_{ik} \\  &  \\  
& =-\frac{4\Lambda }{D-2}\bar G_{ik}\exp {\left[ -\frac
2{D_0-2}\sum_{i=1}^nD_i\beta _c^i\right] ,} 
\end{array}
\end{equation}
so that the auxiliary matrix $\bar G^{-1}A_c$ is proportional to the $n-$%
dimensional identity matrix $I_n$ 
\begin{equation}
\label{4.3}\bar G^{-1}A_c=-\frac{4\Lambda _{eff}}{D_0-2}I_n\ 
\end{equation}
and exciton masses $m_i^2$, in the previous section defined as eigenvalues
of $\bar A_c$ or $\bar G^{-1}A_c$, are simply given as 
\begin{equation}
\label{4.4} 
\begin{array}{ll}
m_1^2=\ldots =m_n^2=m^2=-\frac{4\Lambda _{eff}}{D_0-2} & =- 
\frac{4\Lambda }{D-2}\exp {\left[ -\frac 2{D_0-2}\sum_{i=1}^nD_i\beta
_c^i\right] } \\  &  \\  
& =2\left| \tilde C\right| ^{\frac{D-2}{D_0-2}}\prod_{i=1}^n\left| \frac{D_i 
}{R_i}\right| ^{\frac{D_i}{D_0-2}}. 
\end{array}
\end{equation}
From (\ref{4.1}) and (\ref{4.4}) we see that massive excitons can only occur
when scalar curvature as well as bare and effective cosmological constant
are negative: $R_k,\Lambda ,\Lambda _{eff}<0$. \ The additional requirement $%
\left| \Lambda _{eff}\right| \leq 10^{-54}{\mbox{cm}}^{-2}\approx
10^{-120}\Lambda _{Pl}$ leads not only to an upper bound for the masses of
excitons $m\leq 10^{-60}M_{Pl}\sim 10^{-55}g\ll M_{Pl}\sim 10^{-5}g$,
independently from the number of scale factors, but also strongly narrows
the class of possible internal space configurations. Let us demonstrate the
latter fact with the help of three models with a different number of scale
factors.

{\bf - a) one-scale-factor model:}

Assuming that for a space-time configuration $M_0\times M_1$ with
four-dimensional external space-time $(D_0=4)$ and compact internal factor
space $M_1=H^{D_1}/\Gamma $ with constant negative curvature 
$R_1=-D_1(D_1-1)$
there exists a minimum of the effective potential at $%
a_c=10^2L_{Pl}$ we get $m^2=2(D_1-1)10^{-2(D_1+2)}M_{Pl}^2$ and $\Lambda
_{eff}=-(D_1-1)10^{-2(D_1+2)}\Lambda _{Pl}.$ Thus, according to (\ref{3.20}%
), $\left| \Lambda _{eff}\right| \leq 10^{-120}\Lambda _{Pl}$, the
dimension of the internal space should be at least $D_1=59$.

{\bf - b) two-scale-factor model:}

Extending the previous example, let us suppose that $D_0=4\ ;\
M_1=H^{D_1}/\Gamma _1:\ R_1=-D_1(D_1-1),\ D_1=2,\ a_{(c)1}=10^2L_{Pl}\ \
;\ M_2=H^{D_2}/\Gamma _2:\ R_2=-D_2(D_2-1).$ Effective cosmological constant
and fine-tuning condition (\ref{4.1}) read in this case: 
\begin{equation}
\label{4.5} 
\begin{array}{l}
\Lambda _{eff}=-(D_2-1)^{-D_2/2}\cdot 10^{-2(D_2+4)}\Lambda _{Pl}, \\ 
a_{(c)2}=(D_2-1)^{1/2}a_{(c)1}=(D_2-1)^{1/2}10^2L_{Pl}. 
\end{array}
\end{equation}
Thus, conditions (\ref{3.19a}) are fulfilled for internal spaces $M_2$ with
dimensions $D_2\geq D_{2,crit}=40$. Indeed, in the case of $D_2=40$ we have 
$\ \Lambda _{eff}\simeq -10^{-120}\Lambda _{Pl},\ a_{(c)2}\simeq 6\cdot
10^2L_{Pl}\ $ and hence for $D_2>40\ $there hold the relations$\quad m_i\ll
M_{Pl},\ \left| \Lambda _{eff}\right| <10^{-120}\Lambda _{Pl},\quad
a_{(c)i}\mbox{ \small $^{>}_{\sim}$ }L_{Pl}\ $ as required in (\ref{3.19a}).

{\bf - c) $n-$scale-factor model:}

For simplicity we assume $D_0=4$ and an internal space consisting of $n$
factor spaces of the same type $M_i=H^{D_i}/\Gamma _i$:\ 
\mbox{$R_i=-D_i(D_i-1)$,}\
$D_i=2,\ a_{(c)i}=10^2L_{Pl}.$ The effective cosmological constant is
then given as 
\begin{equation}
\label{4.6}\Lambda _{eff}=-10^{-4(n+1)}\Lambda _{Pl}, 
\end{equation}
so that at least $n=29$ spaces $M_i=H^2/\Gamma _i$ are necessary to fulfill
condition $\left| \Lambda _{eff}\right| <10^{-120}\Lambda _{Pl}.$ 

Summarizing the three examples we can say that for an effective potential of
pure geometrical type, according to observational data, gravitational
excitons should be extremely light particles with masses $m\leq 10^{-55}g$
caused by inhomogeneous scale factor fluctuations of a composite internal
factor space with negative curvature and sufficiently high dimension greater
than some critical dimension. The value of this critical dimension depends
on the topological structure of the internal factor space.

As conclusion we note that a conformal transformation $g^{(1)}\rightarrow
D_1^2g^{(1)}$ with fixed $\kappa _0^2=\kappa ^2/\mu $, $\Lambda =2R_1$ and $%
R_1$ leads in the limit $D_1\rightarrow \infty $ to $a_c\rightarrow L_{Pl}$
and $\Lambda _{eff}\rightarrow 0$. But at the same time the exciton mass
vanishes $(m\rightarrow 0)$ and the effective potential degenerates into a
step function with infinite height: $U_{eff}\rightarrow \infty $ for $a<1$
and $U_{eff}=0$ for $a\geq 1$. Thus, in the limit $D_1\rightarrow \infty $
there is no minimum at all. To satisfy the strong condition $\Lambda
_{eff}=0 $ we should consider the case $\rho \not \equiv 0$.

In the next sections we analyze three concrete types of nonvanishing
potentials $\rho $, originating in the presence of additional fields ---
Casimir potential, perfect fluid potential and ''monopole'' potential.

\section{Casimir and Casimir-like \newline potentials $\rho $\label{mark3}}

\setcounter{equation}{0}

Because of a nontrivial topology of the space - time, vacuum fluctuations of
quantized fields result in a non-zero energy density.

{\bf - a) One-scale-factor model:}

For internal spaces with only one scale factor this energy density has the
form \cite{6,9,12,32,32a,33} 
\begin{equation}
\label{5.1}\rho =Ce^{-D\beta }, 
\end{equation}
where $C$ is a constant that strongly depends on the topology of the model.
For example, for fluctuations of scalar fields the constant $C$ was
calculated to take the values: $C=-8.047\cdot 10^{-6}$ if \mbox{$M_0=\RR \times
S^3$,}\ \mbox{$M_1=S^1$} (with $e^{\beta ^0}$ as scale factor of $S^3$ and $e^{\beta
^0}\gg e^{\beta ^1}$) \cite{9}; \mbox{$C=-1.097$} if $M_0=\RR \times \RR ^2,\
M_1=S^1$ \cite{32} and $C=3.834\cdot 10^{-6}$ if \mbox{$M_0=\RR \times S^3$,}\ 
$M_1=S^3$ (with $e^{\beta ^0}\gg e^{\beta ^1}$) \cite{9}.

For an effective potential with $\rho -$term (\ref{5.1}) (and $n=1$) the
zero-extremum-conditions $\left. \frac{\partial U_{eff}}{\partial \beta }%
\right| _{min}=0$ and $\Lambda _{eff}=0$ lead to a fine tuning of the
parameters of the model 
\begin{equation}
\label{5.2}R_1e^{-2\beta _c}=\frac{2D}{D-2}\Lambda ,\quad R_1e^{(D-2)\beta
_c}=\kappa ^2CD\ 
\end{equation}
which implies $\mbox{\rm sign}\,R_1=\mbox{\rm sign}\,\Lambda =\mbox{\rm sign}%
\,C$. We note that a similar fine tuning was obtained by different methods
in papers \cite{12} (for one internal space) and \cite{20} (for $n$
identical internal spaces).

The second derivative and mass squared read respectively 
\begin{eqnarray}\label{5.3}
a_{11} & = & \left.\frac{\partial ^2U_{eff}}{\partial \beta
^2}\right|_{\beta _c} = (D-2)R_1
{\left(e^{-2\beta _c}\right)}^{\frac{D-2}{D_0-2}}, \\
m^2 & = & \frac{D_0-2}{D_1}R_1
{\left(e^{-2\beta _c}\right)}^{\frac{D-2}{D_0-2}}.
\end{eqnarray}  Thus, the internal space should have positive curvature: $%
R_1>0$ (or for split space $M_1$ the sum of the curvatures of the
constituent spaces $M_1^k$ should be positive).

Let us now perform an explicit test for a manifold $M$ with \mbox{topology}
$M=\RR \times S^3\times S^3$, 
where $e^{\beta ^0}\gg e^{\beta ^1}$. Then \cite{9} 
\mbox{$C=3.834\cdot 10^{-6}>0$} 
and, as $C,R_1>0$, the effective potential has a
minimum provided $\Lambda >0$. Normalizing $\kappa _0^2$ to unity, we get $%
\kappa ^2=\mu $ where $\mu =\frac{2\pi ^{(d+1)/2}}{\Gamma \left(
\frac 12(d+1)\right) }$ is the volume of the $d$ - dimensional sphere. So we
obtain $a_c\approx 1.5\cdot 10^{-1}L_{Pl}$ and $m\approx 2.12\cdot
10^2M_{Pl} $, and conditions (\ref{3.19a}) (i) and (ii) are not satisfied
for this topology. For other topologies this problem needs a separate
investigation.

{\bf - b) Two-scale-factor models:}

After these brief considerations on Casimir potentials for one-scale-factor
models we turn now to some methods applicable for an analysis of
two-scale-factor models with Casimir-like potentials. We proposed the use of
such potentials of the general form 
\begin{equation}
\label{5.4}
\begin{array}{ll}
\rho = & e^{-\sum_{i=1}^nD_i\beta ^i}\sum\Sb k_1,\ldots ,k_n=1 \\
k_2<\cdots <k_n\endSb ^n\left| \epsilon _{k_1k_2\ldots k_n}\right| \sum_{\xi
_1=0}^{D_{k_2}}\ldots \\ 
 & \\ 
 & \ldots \sum_{\xi _{n-1}=0}^{D_{k_n}}A_{\xi _1\ldots \xi
_{n-1}}^{(k_1)}\frac{{\left( e^{\beta ^{k_2}}\right) }^{\xi _1}\ldots {%
\left( e^{\beta ^{k_n}}\right) }^{\xi _{n-1}}}{{\left( e^{\beta
^{k_1}}\right) }^{D_0+\xi _1+\cdots +\xi _{n-1}}} 
\end{array}
\end{equation}
in our paper \cite{22} in order to achieve a first crude insight into a
possible stabilization mechanism of internal space configurations due to
exact Casimir potentials depending on $n$ scale factors. (In (\ref{5.4}) $%
\epsilon _{ik\ldots m}$ denotes the totally antisymmetric symbol 
($\epsilon _{12\ldots n}=+1$) and $A_{\xi _0\ldots \xi
_{n-1}}^{(k_0)} $ are dimensionless constants which depend on the topology
of the model.)

From investigations performed in the last decades (see e.g. \cite{32a,33}
and Refs. therein) we know that exact Casimir potentials can be expressed in
terms of Epstein zeta function series with scale factors as parameters.
Unfortunately, the existing integral representations of these zeta function
series are not well suited for a stability analysis of the effective
potential $U_{eff}$ as function over the total target space $\vec \beta \in $
$\RR
_T^n$. The problems can be circumvented partially by the use of asymptotic
expansions of the zeta function series in terms of elementary functions for
special subdomains $\Omega _a$ of the target space $\Omega _a$ $\subset \RR
_T^n$. According to \cite{32a,33} potential (\ref{5.4}) gives a crude
approximation of exact Casimir potentials in subdomains $\Omega _a$. In
contrast with other approximative potentials proposed in literature \cite
{11,18} potential (\ref{5.4}) shows a physically correct behavior under
decompactification of factor space components \cite{22}. The question, in as
far (\ref{5.4}) can be used in regions $\RR
_T^n\setminus \Omega _a$, needs an additional investigation. The philosophy
of the proposed method consists in a consideration of potentials (\ref{5.4})
on the whole target space $\RR
_T^n$, and testing of scale factors and parameters of possible minima of
the corresponding effective potential on their compatibility with asymptotic
approximations of exact Casimir potentials in $\Omega _a$. As a beginning,
we describe in the following only some techniques, without explicit
calculation and estimation of exciton masses.

Before we start our analysis of two-scale-factor models with Casimir-like
potentials 
\begin{equation}
\label{5.5}\rho =e^{-\sum_{i=1}^2D_i\beta ^i}\left[
\sum_{i=0}^{D_2}A_i^{(1)} \frac{{e^{i\beta ^2}}}{{e^{(D_0+i)\beta ^1}}}%
+\sum_{j=0}^{D_1}A_j^{(2)}\frac{{e^{j\beta ^1}}}{{e^{{(D_0+j)}\beta ^2}}}%
\right] 
\end{equation}
let us introduce the following convenient (temporary) notations: \\
\mbox{$x:=a_1\equiv \ {e^{\beta ^1}}$,}\ 
$y:=a_2\equiv {e^{\beta ^2}},\ P_\xi :=\kappa
^2A_\xi ^{(1)},\ S_\xi :=\kappa ^2A_\xi ^{(2)}.\ $In terms of these
notations the effective potential (\ref{2.12}) reads 
\begin{equation}
\label{5.6} 
\begin{array}{ll}
U_{eff}= & (x^{D_1}y^{D_2})^{-\frac 2{D_0-2}}\left[ - 
\frac{R_1}2x^{-2}-\frac{R_2}2y^{-2}+\Lambda +\right. \\  &  \\  
& \left. +x^{-D_1}y^{-D_2}\left(
\sum_{i=0}^{D_2}P_iy^ix^{-(D_0+i)}+\sum_{j=0}^{D_1}S_jx^jy^{-(D_0+j)}\right)
\right] .
\end{array}
\end{equation}
For physically relevant configurations with scale-factors near Planck length 
\begin{equation}
\label{5.7}0<x,y<\infty 
\end{equation}
we transform extremum conditions $\partial _{\beta ^{1,2}}U_{eff}=0
\Leftrightarrow \partial _{x,y}U_{eff}=0$ by factoring out of $(xy)^{-D}-$%
terms and taking combinations \\
\mbox{$\partial _xU_{eff}\pm \partial _yU_{eff}=0$}
to an equivalent system of two algebraic equations in $x$ and $y$ : 
%\vspace{-1ex}
\begin{equation}
\label{5.8} 
\begin{array}{lll}
I_{1+} & = & (xy)^{D-2}\left[ 
\frac{D-2}{D_0-2}\left( R_1y^2+R_2x^2\right) -\frac{2\Lambda }{D_0-2}%
D^{^{\prime }}x^2y^2\right] - \\[-1ex]  &  &  \\[-1ex]  
&  & -\left( 
\frac{2D^{^{\prime }}}{D_0-2}+D\right) \left[
\sum_{i=0}^{D_2}P_iy^{D_0+D_1+i}x^{D_2-i}+\right. \\[-1ex]  &  &  \\[-1ex]  
&  & \left. +\sum_{j=0}^{D_1}S_jy^{D_1-j}x^{D_0+D_2+j}\right] \\[-1ex]  
&  &  \\[-1ex]  
& = & 0 \\
 & \\
I_{1-} & = & (xy)^{D-2}\left[ 
\frac{D_1-D_2}{D_0-2}\left( R_1y^2+R_2x^2\right) +\left(
R_1y^2-R_2x^2\right) -\right. \\[-1ex]  &  &  \\[-1ex]  
&  & \left. - 
\frac{2\Lambda }{D_0-2}\left( D_1-D_2\right) x^2y^2\right] - \\[-1ex]
  &  &  \\[-1ex]  
&  & -\sum_{i=0}^{D_2}P_i\left[ D_0\left( 
\frac{D_1-D_2}{D_0-2}+1\right) +2i\right] y^{D_0+D_1+i}x^{D_2-i}- \\[-1ex] 
 &  &  
\\[-1ex]  
&  & -\sum_{j=0}^{D_1}S_j\left[ D_0\left( 
\frac{D_1-D_2}{D_0-2}-1\right) -2j\right] y^{D_1-j}x^{D_0+D_2+j} \\[-1ex]  &  &  
\\[-1ex]  
& = & 0. 
\end{array}
\end{equation}
Thus, scale-factors $a_1$ and $a_2$ satisfying the extremum conditions are
defined as common roots of polynomials (\ref{5.8}).
%, (\ref{5.8a}). 
In the general case of
arbitrary dimensions $(D_0,\ D_1,\ D_2)$ and arbitrary parameters 
$\left\{ R_1,R_2,\right. $\\ $\left. P_i,S_i\right\} $ 
these roots are complex, so that only a restricted
subclass of them are real and fulfill condition (\ref{5.7}). In the
following we derive necessary conditions on the parameter set guaranteeing
the existence of real roots satisfying (\ref{5.7}). The analysis could be
carried out using resultant techniques \cite{34} on variables $x,\ y$
directly. The structure of $I_{1\pm }$ suggests another, more convenient
method \cite{35}. Introducing the projective coordinate $\lambda =y/x$ we
rewrite (\ref{5.8})
%, (\ref{5.8a}) 
as $I_{1\pm }=x^DI_{2\pm }(y,\lambda )$ with 
\begin{equation}
\label{5.9} 
\begin{array}{lllccccc}
I_{2+} & = & -a_0(\lambda )+a_{D-2}(\lambda )y^{D-2}-y^D\Delta _{+} & = & 0
&  &  & (a) \\  
&  &  &  &  &  &  &  \\ 
I_{2-} & = & -b_0(\lambda )+b_{D-2}(\lambda )y^{D-2}-y^D\Delta _{-} & = & 0
&  &  & (b) 
\end{array}
\end{equation}
and coefficient-functions 
\begin{equation}
\label{5.10} 
\begin{array}{lll}
a_0(\lambda ) & = & [ 
\frac{2D^{^{\prime }}}{D_0-2}+D]\left[ \sum_{i=0}^{D_2}P_i\lambda
^{D_0+D_1+i}+\sum_{j=0}^{D_1}S_j\lambda ^{D_1-j}\right] \\  &  &  \\ 
a_{D-2}(\lambda ) & = & \frac{D-2}{D_0-2}\left( R_1\lambda ^2+R_2\right) \\  
&  &  \\ 
b_0(\lambda ) & = & \sum_{i=0}^{D_2}P_i\left[ D_0\left( 
\frac{D_1-D_2}{D_0-2}+1\right) +2i\right] \lambda ^{D_0+D_1+i}+ \\  &  &  \\
&  & +\sum_{j=0}^{D_1}S_j\left[ D_0\left( 
\frac{D_1-D_2}{D_0-2}-1\right) -2j\right] \lambda ^{D_1-j} \\  &  &  \\ 
b_{D-2}(\lambda ) & = & \frac{D_1-D_2}{D_0-2}\left( R_1\lambda ^2+R_2\right)
+\left( R_1\lambda ^2-R_2\right) \\  &  &  \\ 
\Delta _{\pm } & = & \frac{2\Lambda }{D_0-2}(D_1\pm D_2). 
\end{array}
\end{equation}
Equations (\ref{5.9}) have common roots if the coefficient functions 
$\{a_i(\lambda ),\\ b_i(\lambda )\}$ 
are connected by a constraint. This
constraint is given by the vanishing resultant 
\begin{equation}
\label{5.11}R_y[I_{2+},I_{2-}]=w(\lambda )=0. 
\end{equation}
Now, the roots can be obtained in two steps. First, one finds the set of
roots $\{\lambda _i\}$ of the polynomial $w(\lambda )$. Physical condition (%
\ref{5.7}) on the affine coordinates $(x,y)$ implies here a corresponding
condition on the projective coordinate $\lambda =y/x$ 
\begin{equation}
\label{5.12}Im(\lambda )=0,\qquad 0<\lambda <\infty . 
\end{equation}
Second, one searches for each $\lambda _i$ solutions $\{y_{ij}\}$ of (\ref
{5.9}). The complete set of physically relevant solutions of system (\ref
{5.8}) 
%(\ref{5.8a}) 
is then given in terms of pairs $\{x_{ij}=y_{ij}/\lambda _i,\
y_{ij}\} $.

Because of the simple $y-$structure of equations (\ref{5.9}) the polynomial $%
w(\lambda )$ can be derived from (\ref{5.9}) directly, without explicit
calculation of the resultant. Taking $b_0(\lambda )I_{2+}-a_0(\lambda
)I_{2-}=0$, \\
\mbox{$\Delta _{-}I_{2+}-\Delta _{+}I_{2-}=0$} 
and assuming $y>0$ we
get 
\begin{equation}
\label{5.13}y^2=\frac{L_3}{L_1},\qquad y^{D-2}=\frac{L_1}{L_2}, 
\end{equation}
where 
\begin{equation}
\label{5.14} 
\begin{array}{lll}
L_1(\lambda ) & := & \Delta _{-}a_0(\lambda )-\Delta _{+}b_0(\lambda ) \\  
&  &  \\ 
L_2(\lambda ) & := & \Delta _{-}a_{D-2}(\lambda )-\Delta _{+}b_{D-2}(\lambda
) \\  
&  &  \\ 
L_3(\lambda ) & := & a_0(\lambda )b_{D-2}(\lambda )-b_0(\lambda
)a_{D-2}(\lambda )\  
\end{array}
\end{equation}
depend only on $\lambda $. Excluding $y$ from (\ref{5.13}) yields the
necessary constraint for the coefficient functions of equation system (\ref
{5.9}) 
\begin{equation}
\label{5.15}w(\lambda )=L_2^2(\lambda )L_3^{D-2}(\lambda )-L_1^D(\lambda )=0. 
\end{equation}
Together with condition (\ref{5.12}), this polynomial of degree 
\begin{equation}
\deg{}_\lambda [w(\lambda )]=D^2
\end{equation}
can be used for a first test of internal space
configurations on stability of their compactification. If the corresponding
parameters $\left\{ R_1,R_2,P_i,S_i\right\} $ allow the existence of
positive real roots $\lambda _i$, the space configuration is a possible
candidate for a stable compactified configuration and can be further tested
on the existence of minima of the effective potential $U_{eff}.$ Otherwise
it belongs to the class of unstable internal space configurations.

Before we turn to the consideration of two-scale-factor models with factor
spaces of the same topological type $(M_1=M_2)$ we note that for the
coefficient functions (\ref{5.14}), because of (\ref{5.7}) and (\ref{5.13}), 
there must hold 
\begin{equation}
\label{5.16}sign(L_1)\big|
_{\lambda _i}=sign(L_2)\big|
_{\lambda _i}=sign(L_3)\big|
_{\lambda _i}. 
\end{equation}
Furthermore we see from (\ref{5.15}) that for even dimensions \\ $D=\dim
(M_1)+\dim (M_2)+\dim (M_0)$ of the product-manifold the polynomial $%
w(\lambda )$ factors into two subpolynomials of degree $D^2/2$%
\begin{equation}
\label{5.17}w(\lambda )=\left[ L_2(\lambda )L_3^{\frac{D-2}2}(\lambda
)+L_1^{\frac D2}(\lambda )\right] \left[ L_2(\lambda )L_3^{\frac{D-2}%
2}(\lambda )-L_1^{\frac D2}(\lambda )\right] =0. 
\end{equation}

\newpage
{\bf - b,1) Two identical internal factor-spaces:}

In the case of identical internal factor-spaces $M_1$ and $M_2$ we have $\ D{%
_1=D_2,\ P_i=S_i,\ R_1=R_2}$. If we assume additionally an external
space-time $M_0$ with $\dim M_0=4$ and, hence, ${\ D=2(D_1+2),\ }$ then
equations (\ref{5.9}) and polynomial (\ref{5.17}) can be rewritten as 
\begin{equation}
\label{5.18} 
\begin{array}{llccc}
I_{2+}= & -4(D_1+1)\bar a_0(\lambda )+(D_1+1)R_1(\lambda ^2+1)y^{D-2}- &  &  
&  \\  
&  &  &  &  \\  
& -2D_1\Lambda y^D=0 &  &  & (a) \\  
&  &  &  &  \\ 
I_{2-}= & (\lambda ^2-1)\left[ -2\bar b_0(\lambda )+R_1y^{D-2}\right] =0 &  
&  & (b) 
\end{array}
\end{equation}
and 
\begin{equation}
\label{5.19} 
\begin{array}{ll}
w(\lambda ) & =4\Lambda ^2D_1^2\left[ 2(\lambda ^2-1)\right]
^{2(D_1+2)}\times \\  
&  \\  
& \times 
\stackunder{w_{+}(\lambda )}{\underbrace{\left[ R_1^{D_1+2}\left(
2(D_1+1)\right) ^{D_1+1}\bar L_3^{D_1+1}+\left( -2\Lambda D_1\right)
^{D_1+1}\bar b_0^{D_1+2}\right] }}\times \\  &  \\  
& \times 
\stackunder{w_{-}(\lambda )}{\underbrace{\left[ R_1^{D_1+2}\left(
2(D_1+1)\right) ^{D_1+1}\bar L_3^{D_1+1}-\left( -2\Lambda D_1\right)
^{D_1+1}\bar b_0^{D_1+2}\right] }} \\  &  \\  
& =0 
\end{array}
\end{equation}
with the notations 
\begin{equation}
\label{5.20} 
\begin{array}{lll}
\bar L_3 & := & 2\bar a_0-(\lambda ^2+1)\bar b_0 \\  
&  &  \\ 
\bar a_0 & := & \sum_{i=0}^{D_1}P_i\left[ \lambda ^{4+D_1+i}+\lambda
^{D_1-i}\right] \\  
&  &  \\ 
\bar b_0 & := & \sum_{i=0}^{D_1}P_i(2+i)\lambda
^{D_1-i}\sum_{j=0}^{i+1}\lambda ^{2j}. 
\end{array}
\end{equation}
From (\ref{5.18}) and (\ref{5.19}) we see that the constraint (\ref{5.11})
is trivially satisfied for coinciding scale-factors $x=y$, i.e. $\lambda =1$.
 Although (\ref{5.19}) holds for all $\lambda $ corresponding to extrema of
the effective potential, roots $y$ can be obtained from relations (\ref{5.13}%
) only for $\lambda \neq 1$. In this nondegenerated case (\ref{5.13}) reads 
\begin{equation}
\label{5.21}y^2=-\frac{(D_1+1)R_1\bar L_3}{2\Lambda D_1\bar b_0},\qquad
y^{D-2}=\frac{2\bar b_0}{R_1}. 
\end{equation}
In the degenerated case $\lambda =1$ relations (\ref{5.13}) become undefined
of type $0/0$ and the scale factor $y$ at the extremum point of the
effective potential (\ref{5.6}) must be found as a root of the polynomial $%
I_{2+}$ (\ref{5.18}(a)) directly. For this polynomial we have now simply 
\begin{equation}
\label{5.22}I_{2+}(\lambda =1):=\frac{\Lambda D_1}{D_1+1}%
y^{2(D_1+2)}-R_1y^{2(D_1+1)}+4\sum_{i=0}^{D_1}P_i=0. 
\end{equation}
Coming back to the general case of identical factor-spaces $M_1,\ M_2$ with
coinciding or noncoinciding scale factors we note that there exists an
interchange symmetry between $M_1$ and $M_2$, which becomes apparent in the
root structure of the polynomial $w(\lambda )$. From 
\begin{equation}
\label{5.23} 
\begin{array}{ll}
U_{eff}= & (xy)^{-D_1}\left[ - 
\frac{R_1}2(x^{-2}-y^{-2})+\Lambda +\right. \\  &  \\  
& \left. +(xy)^{-D_1}\sum_{i=0}^{D_1}P_i\left(
y^ix^{-4-i}+x^iy^{-4-i}\right) \right] 
\end{array}
\end{equation}
we see that $x$ and $y$ enter (\ref{5.23}) symmetrically. When one extremum
of (\ref{5.23}) is located at $\{x_i=a,\ y_i=b\}$ then because of the
interchange symmetry $x\rightleftharpoons y$ there exists a second extremum
located at 
$\{x_j=b,\\ y_j=a\}$. 
So we have for the corresponding projective
coordinates : 
\begin{equation}
\label{5.24}\lambda _i=y_i/x_i=b/a,\quad \lambda _j=y_j/x_j=a/b\
\Longrightarrow \lambda _i=\lambda _j^{-1}. 
\end{equation}
By regrouping of terms in (\ref{5.19}) it is easy to show that 
\begin{equation}
\label{5.25}w(\lambda ^{-1})=\lambda ^{-D^2}w(\lambda ) 
\end{equation}
and, hence, roots $\{\lambda _i\neq 0\}$ of $w(\lambda )=0$ exist indeed in
pairs $\{\lambda _i,\ \lambda _i^{-1}\}.$ But there is no relation
connecting this root-structure with a symmetry between $w_{+}(\lambda )$ and 
$w_{-}(\lambda )$ in (\ref{5.19}) $w_{+}(\lambda ^{-1})\not \sim
w_{-}(\lambda )$. For completeness, we note that relation (\ref{5.24}) is
formally similar to dualities recently investigated in superstring theory 
\cite{36a}.

Before we turn to an analysis of minimum conditions for effective potentials 
$U_{eff}$ corresponding to special classes of solutions of $w(\lambda )=0$
we rewrite the necessary second derivatives 
\begin{equation}
\label{5.25a} 
\begin{array}{ll}
\partial _{xx}^2U_{eff}= & - 
\frac{R_1}2\left( \alpha _1x^{-D_1-4}y^{-D_1}+\alpha
_2x^{-D_1-2}y^{-D_1-2}\right) + \\  &  \\  
& +\Lambda \alpha _2x^{-D_1-2}y^{-D_1}+ \\  
&  \\  
& +\sum_{i=0}^{D_1}P_i\left( \alpha _3y^{i-2D_1}x^{-i-2D_1-6}+\alpha
_4x^{i-2D_1-2}y^{-i-2D_1-4}\right) \\  
&  \\ 
\partial _{yy}^2U_{eff}= & \partial _{xx}^2U_{eff} 
\Big|_{x\rightleftharpoons y} \\  &  \\ 
\partial _{xy}^2U_{eff}= & - 
\frac{R_1}2\alpha _5\left( x^{-D_1-3}y^{-D_1-1}+x^{-D_1-1}y^{-D_1-3}\right)
+ \\  &  \\  
& +\Lambda \alpha _6x^{-D_1-1}y^{-D_1-1}+\sum_{i=0}^{D_1}P_i\alpha _7\left(
y^{i-2D_1-1}x^{-i-2D_1-5}+\right. \\  
&  \\  
& \left. +\alpha _4x^{i-2D_1-1}y^{-i-2D_1-5}\right) , 
\end{array}
\end{equation}
where 
\begin{equation}
\label{5.25b} 
\begin{array}{lllll}
\alpha _1= & (D_1+2)(D_1+3) &  & \alpha _2= & D_1(D_1+1) \\  
&  &  &  &  \\ 
\alpha _3= & (2D_1+i+4)(2D_1+i+5) &  & \alpha _5= & D_1(D_1+2) \\  
&  &  &  &  \\ 
\alpha _4= & (2D_1-i)(2D_1-i+1) &  & \alpha _6= & D_1^2 \\  
&  &  &  &  \\ 
\alpha _7= & (2D_1+i+4)(2D_1-i), &  &  &  
\end{array}
\end{equation}
in the more appropriate form (notation $\widetilde{\mu }=\lambda
^{D_1}y^{-2D+2}$) 
\begin{equation}
\label{5.25c} 
\begin{array}{ll}
\partial _{xx}^2U_{eff}= & \lambda ^2 
\widetilde{\mu }\left[ -\frac{R_1}2\left( \alpha _1\lambda ^2+\alpha
_2\right) y^{D-2}+\Lambda \alpha _2y^D+\right. \\  &  \\  
& \left. +\sum_{i=0}^{D_1}P_i\left( \alpha _3\lambda ^{4+D_1+i}+\alpha
_4\lambda ^{D_1-i}\right) \right] \\  
&  \\ 
\partial _{yy}^2U_{eff}= & \widetilde{\mu }\left[ -\frac{R_1}2\left( \alpha
_1+\alpha _2\lambda ^2\right) y^{D-2}+\Lambda \alpha _2y^D+\right. \\  &  \\
& \left. +\sum_{i=0}^{D_1}P_i\left( \alpha _4\lambda ^{4+D_1+i}+\alpha
_3\lambda ^{D_1-i}\right) \right] \\  
&  \\ 
\partial _{xy}^2U_{eff}= & \lambda 
\widetilde{\mu }\left[ -\frac{R_1}2\alpha _5\left( \lambda ^2+1\right)
y^{D-2}+\Lambda \alpha _6y^D+\right. \\  &  \\  
& \left. +\sum_{i=0}^{D_1}P_i\alpha _7\left( \lambda ^{4+D_1+i}+\lambda
^{D_1-i}\right) \right] . 
\end{array}
\end{equation}
Introducing the notations 
\begin{equation}
\label{5.25c001}\widetilde{A}_c:=\left( 
\begin{array}{cc}
\partial _{xx}^2U_{eff} & \partial _{xy}^2U_{eff} \\ 
\partial _{xy}^2U_{eff} & \partial _{yy}^2U_{eff} 
\end{array}
\right) 
\end{equation}
and 
\begin{equation}
\label{5.25c002}w_{(c)1,2}:=\frac 12\left[ Tr(\widetilde{A}_c)\pm \sqrt{%
Tr^2( \widetilde{A}_c)-4\det (\widetilde{A}_c)}\right] 
\end{equation}
the minimum conditions are given as 
\begin{equation}
\label{5.25c003}w_{(c)1}>0,\ w_{(c)2}\geq 0. 
\end{equation}
In the degenerated case of coinciding scale-factors $x=y,\ \lambda =1$
there hold the following relations between the derivatives of effective
potentials $U_{eff}(x,y)$ and $\widetilde{U}_{eff}(y)=U_{eff}(y,y)$%
\begin{equation}
\label{5.25c01} 
\begin{array}{l}
\partial _y 
\widetilde{U}_{eff}=\partial _xU_{eff}\big| _{x=y}+\partial _yU_{eff}\big| %
_{x=y} \\  \\ 
\partial _{yy}^2\widetilde{U}_{eff}=\partial _{xx}^2U_{eff}\big| %
_{x=y}+\partial _{yy}^2U_{eff}\big| _{x=y}+2\partial _{xy}^2U_{eff}\big| %
_{x=y} 
\end{array}
\end{equation}
and minimum conditions reduce to 
\begin{equation}
\label{5.25c02}\partial _y\widetilde{U}_{eff}=0,\quad \partial _{yy}^2 
\widetilde{U}_{eff}>0 
\end{equation}
with 
\begin{equation}
\label{5.25c03} 
\begin{array}{ll}
\partial _{yy}^2\widetilde{U}_{eff}= & 2y^{-2D+2}\left[
-R_1(D_1+1)(2D_1+3)y^{D-2}+\right. \\  
&  \\  
& \left. +\Lambda D_1(2D_1+1)y^D+4(D_1+1)(4D_1+5)\sum_{i=0}^{D_1}P_i\right]
. 
\end{array}
\end{equation}
For convenience of the additional explicit calculations of constraint $%
U_{eff}\big|
_{\min }=0$ we rewrite also effective potential (\ref{5.23}) in terms of
variables $y,\ \lambda $%
\begin{equation}
\label{5.25c04} 
\begin{array}{ll}
U_{eff}= & \lambda ^{D_1}y^{-2D+4}\left[ - 
\frac{R_1}2y^{D-2}(\lambda ^2+1)+\Lambda y^D+\right. \\  &  \\  
& \left. +\sum_{i=0}^{D_1}P_i\left( \lambda ^{4+D_1+i}+\lambda
^{D_1-i}\right) \right] . 
\end{array}
\end{equation}
The further analysis consists in a compatibility consideration of minimum
conditions (\ref{5.25c003}) and (\ref{5.25c02}) with properties of the
polynomial $w(\lambda )$, expressions like (\ref{5.21}) defining $y^{D-2}$
and $y^D=y^{D-2}y^2$ as functions of $\lambda $ on the parameter-space 
\mbox{$\RR _{par}^{D_1+3}=\{(R_1,\Lambda ,P_i)\mid i=0,\ldots ,D_1\}$} 
and the
constraint $U_{eff}\big|
_{\min }=0$. As result we will get a first crude division of $\RR %
_{par}^{D_1+3}$ in stability-domains allowing the existence of minima of the
effective potential $U_{eff}$ and forbidden regions corresponding to
instable internal space configurations.

After these general considerations we turn now to a more concrete analysis.

{\bf - b,2) Noncoinciding scale-factors $(\lambda \neq 1),R_1,\Lambda \neq 0$%
:}

First we consider the polynomial $w(\lambda )$. We know that stable
internal space-configurations correspond to real projective coordinates $%
0<\lambda <\infty $. So we have to test subpolynomials $w_{\pm }(\lambda )$
on the existence of such roots. The high degree 
\mbox{$\deg {}_\lambda [w_{\pm}(\lambda )]=2(D_1+2)(D_1+1)\geq 24$,}
$D_1\geq 2$ (because of nonvanishing
curvature of the factor-spaces $M_1,M_2$) allows only an analysis by
techniques of number theory \cite{36}, 
the theory of ideals of commutative rings \cite
{34} or, for general parameter-configurations, numerical tests. In the
latter case the number of effective test-pa\-ra\-meters can be reduced by
introduction of new coordinates in parameter-space 
\begin{equation}
\label{5.25c1} 
\begin{array}{ll}
\RR _{par}^{D_1+3}\rightarrow \overline{\RR }_{par}^{D_1+1}= & \left\{ (\chi
,p_i)\mid \chi =\left( 
\frac{2\Lambda D_1}{D_1+1}\right) ^{D_1+1}\frac{2P_0}{R_1^{D_1+2}},\right.
\\  &  \\  
& \left. p_i=\frac{P_i}{P_0},\ i=1,\ldots ,D_1\right\} 
\end{array}
\end{equation}
(for $P_0\neq 0,\ p_0=1$ ; in the opposite case $P_0$ can be replaced by
any nonzero $P_i$). Polynomials $w_{\pm }(\lambda )=0$ transform then to $%
w_{\pm }(\lambda )=\frac 12R_1^{D_1+2}\left( (D_1+1)P_0\right) ^{D_1+1}\bar
w_{\pm }(\lambda )=0$, where 
\begin{equation}
\label{5.25c2} 
\begin{array}{l}
\bar w_{\pm }(\lambda ):= 
\widetilde{L}_3^{D_1+1}\pm (-)^{D_1+1}\chi \widetilde{b}_0^{D_1+2}=0 \\  \\ 
\widetilde{L}_3:=\bar L_3(P_0=1;P_1=p_1,\ldots ,P_{D_1}=p_{D_1}) \\  \\ 
\widetilde{b}_0:=\bar b_0(P_0=1;P_1=p_1,\ldots ,P_{D_1}=p_{D_1}). 
\end{array}
\end{equation}
Test are easy to perform with programs like {\sc mathematica}
%\copyright \ 
or 
{\sc maple}.
%\copyright .

As a second step we have to consider minimum conditions (\ref{5.25c003}).
Using (\ref{5.21}) we substitute 
\begin{equation}
\label{5.25d}y^{D-2}=\frac{2\bar b_0}{R_1}\ ;\qquad y^D=-\frac{(D_1+1)\bar
L_3}{\Lambda D_1} 
\end{equation}
into (\ref{5.25c}) and transform (\ref{5.25c003}) to the following
equivalent inequalities 
\begin{equation}
\label{5.25e} 
\begin{array}{l}
2(D_1+1)(D_1+2)Q_1(\lambda )+Q_2(\lambda )>(D_1+2)(\lambda ^2+1)\bar
b_0(\lambda ) \\  
\\ 
[0ex][2(D_1+2)Q_1(\lambda )-(\lambda ^2+1)\bar b_0(\lambda )][Q_2(\lambda
)-(\lambda ^2+1)\bar b_0(\lambda )]\geq \\ 
\geq (D_1+1)(\lambda ^2-1)^2\bar b_0^2(\lambda ) 
\end{array}
\end{equation}
with notations 
\begin{equation}
\label{5.25f} 
\begin{array}{l}
Q_1(\lambda ):=\sum_{i=0}^{D_1}P_i\left( \lambda ^{4+D_1+i}+\lambda
^{D_1-i}\right) \equiv \bar a_0(\lambda ) \\  
\\ 
Q_2(\lambda ):=\sum_{i=0}^{D_1}P_i(i+2)^2\left( \lambda ^{4+D_1+i}+\lambda
^{D_1-i}\right) . 
\end{array}
\end{equation}

Stability-domains in parameter-space $\RR _{par}^{D_1+3}$, corresponding to
minima of the effective potential are given as intersections of domains
defined by (\ref{5.25e}) with domains which allow the existence of physical
relevant roots of $\bar w_{\pm }(\lambda )=0$. So numerical tests on minima
are easy to perform. If we additionally assume that $U_{eff}\big|
_{\min }=0$ then the class of possible stability domains narrows
considerably. Substitution of (\ref{5.25d}) into (\ref{5.25c04}) transforms
this constraint to 
\begin{equation}
\label{5.25f1}(\lambda ^2+1)\bar b_0(\lambda )=(D_1+2)Q_1(\lambda ) 
\end{equation}
and inequalities (\ref{5.25e}) to 
\begin{equation}
\label{5.25f2} 
\begin{array}{l}
D_1(\lambda ^2+1)\bar b_0(\lambda )+Q_2(\lambda )>0 \\  
\\ 
(\lambda ^2+1)[Q_2(\lambda )-(\lambda ^2+1)\bar b_0(\lambda )]\geq
(D_1+1)(\lambda ^2-1)^2\bar b_0(\lambda )\geq 0. 
\end{array}
\end{equation}

From (\ref{5.25e}) we get additional analytical insight into the minimum
structure of the effective potential. Taking the limit $\lambda \rightarrow
1 $ we have 
\begin{equation}
\label{5.25f3} 
\begin{array}{l}
Q_1(1)=2\sum_{i=0}^{D_1}P_i \\  
\\ 
Q_2(1)=[(\lambda ^2+1)\bar b_0(\lambda )]_{\lambda
=1}=2\sum_{i=0}^{D_1}P_i(i+2)^2 
\end{array}
\end{equation}
so that inequalities (\ref{5.25e}) reduce to 
\begin{equation}
\label{5.25g}2(D_1+2)\sum_{i=0}^{D_1}P_i>\sum_{i=0}^{D_1}P_i(i+2)^2 
\end{equation}
and for $U_{eff}\big|
_{\min }=0$ even to 
\begin{equation}
\label{5.25h}(D_1+2)\sum_{i=0}^{D_1}P_i=\sum_{i=0}^{D_1}P_i(i+2)^2>0. 
\end{equation}
From (\ref{5.25c}), (\ref{5.25c002}) and (\ref{5.25f3}) it is easy to see
that in the case $\lambda =1$ the eigenvalues $w_{(c)1,2}$ of the Hessian $
\widetilde{A}_c$ (\ref{5.25c001}) are given as 
\begin{equation}
\label{5.25h1}w_{(c)1}=(D_1+1)\left[ 2(D_1+2)Q_1(1)-Q_2(1)\right] >0,\quad
w_{(c)2}=0 
\end{equation}
and the minimum of the effective potential in quadratic approximation (\ref
{3.1}) becomes degenerated.

{\bf - b,3) Coinciding scale-factors $(\lambda =1),R_1,\Lambda \neq 0$:}

In this case extrema of the effective potential (\ref{5.23}) are given by
the roots of polynomial $I_{2+}(\lambda =1)$ (\ref{5.22}). From the
structure of $I_{2+}(\lambda =1)$ immediately follows:

1. Because $I_{2+}(\lambda =1)$ contains only terms with even degree in $y$,
 there exist no real roots --- and hence no extrema of the effective
potential $U_{eff}$ --- for parameter combinations with: 
\begin{equation}
\label{5.26}sign\left( \sum_{i=0}^{D_1}P_i\right) =sign(\Lambda )\neq
sign(R_1). 
\end{equation}

2. For arbitrary parameters $\Lambda ,R_1,\bar \Delta
:=\sum_{i=0}^{D_1}P_i$ roots of 
\mbox{$I_{2+}(\lambda =1)$}
can be found by
analytical methods up to dimensions $D_1\leq 2$ performing a substitution $%
z:=y^2$ and using standard techniques for polynomials of degree $\deg
{}_zI_{2+}(\lambda =1)\leq 4$. Because of $R_1\neq 0\ \Leftrightarrow
D_1\geq 2$ such considerations are restricted to the case $D_1=2$.

3. There exist no general mathematical methods to obtain roots of
polynomials with degree $\deg {}_zI_{2+}(\lambda =1)>4$ and {\em arbitrary}
coefficients analytically. For special restricted classes of coefficients
techniques of number theory \cite{36}, are applicable. We do not use such
techniques in the present paper. For polynomials $I_{2+}(\lambda =1)$ and
dimensions $\dim M_1=\dim M_2=D_1>2$ this implies that arbitrary parameter
sets should be analyzed numerically or parameters $\Lambda ,\ R_1,\ $ $\bar
\Delta $ should be fine tuned --- chosen ad hoc in such a way that $%
I_{2+}(\lambda =1)=0$ is fulfilled.

In the following we derive a necessary condition for the existence of a
minimum of the effective potential with fine-tuned parameters. Using the
ansatz 
\begin{equation}
\label{5.27}\frac{\Lambda D_1}{D_1+1}=\sigma _1y_0^{-2}\ ;\quad \bar \Delta
:=\sum_{i=0}^{D_1}P_i=\sigma _2y_0^{D-2} 
\end{equation}
equation (\ref{5.22}) reduces to 
\begin{equation}
\label{5.28}(\sigma _1-R_1+4\sigma _2)y_0^{D-2}=0. 
\end{equation}
Without loss of generality we choose $\sigma _2$ as free parameter, and
hence $\sigma _1=R_1-4\sigma _2$, so that from relations (\ref{5.27}) 
\begin{equation}
\label{5.28a}y_0^{D-2}=\frac{\bar \Delta }{\sigma _2},\quad y_0^D=\frac{%
D_1+1}{\Lambda D_1}\bar \Delta (\frac{R_1}{\sigma _2}-4) 
\end{equation}
and (\ref{5.25c03}) minimum condition (\ref{5.25c02}) reads 
\begin{equation}
\label{5.28b}\left. \partial _{yy}^2\widetilde{U}_{eff}\right| _{\min
}=4y_0^{-2D+2}(D_1+1)\left[ 4(D_1+2)-\frac{R_1}{\sigma _2}\right] \bar
\Delta >0 
\end{equation}
or 
\begin{equation}
\label{5.28c}(2D-\frac{R_1}{\sigma _2})\sum_{i=0}^{D_1}P_i>0. 
\end{equation}
We see that there exists a critical value $\sigma _c=\frac{R_1}{2D}$ which
separates stability-domains with different signs of $\bar \Delta $%
\begin{equation}
\label{5.28d} 
\begin{array}{lll}
\bar \Delta =\sum_{i=0}^{D_1}P_i>0 & \Longleftrightarrow & \left| \sigma
_2\right| >\left| \sigma _c\right| \\  
&  &  \\ 
\bar \Delta =\sum_{i=0}^{D_1}P_i<0 & \Longleftrightarrow & \left| \sigma
_2\right| <\left| \sigma _c\right| . 
\end{array}
\end{equation}
To complete our considerations of the degenerated case $(\lambda
=1),R_1$,\\ \mbox{$\Lambda \neq 0$} we derive the constraint $U_{eff}\big|
_{\min }=0$. By use of (\ref{5.25c04}) and (\ref{5.28a}) this is easily
done to yield $\sigma _2=R_1/D=2\sigma _c$. So the constraint fixes the
free parameter $\sigma _2$. Remembering that according to our temporary
notation $y:=a_2\equiv {e^{\beta ^2}\ }$ the value $y_0$ defines the scale
factor of the internal spaces at the minimum position of the effective
potential, we get now for the fine-tuning conditions (\ref{5.27}) 
\begin{equation}
\label{5.29}\Lambda =\frac{(D-2)R_1}{Da_{(c)2}^2},\quad \bar \Delta =\frac{%
R_1a_{(c)2}^{D-2}}D,\quad \bar \Delta ^2=\frac{R_1^D(D-2)^{D-2}}{%
D^D\Lambda ^{D-2}} 
\end{equation}
--- the well-known conditions widely used in literature \cite{20}. From (\ref
{5.28d}), (\ref{5.29}) and $a_{(c)2}>0$ we see that for $\sigma
_2=R_1/D=2\sigma _c$ the stability-domain in parameter-space $\RR %
_{par}^{m+3}$ is narrowed to the sector 
\begin{equation}
\label{5.30}\bar \Delta =\sum_{i=0}^{D_1}P_i>0,\quad R_1>0,\quad \Lambda
>0. 
\end{equation}
\newpage
{\bf - b,4) Vanishing curvature-scalars $(R_1=0),\ \Lambda \neq 0$:}

For vanishing curvature scalars equations (\ref{5.18}) reduce to 
\begin{equation}
\label{5.35} 
\begin{array}{llllllll}
I_{2+} & = & -2(D-2)\bar a_0(\lambda )-(D-4)\Lambda y^D & = & 0 &  &  & (a)
\\  
&  &  &  &  &  &  &  \\ 
I_{2-} & = & -2(\lambda ^2-1)\bar b_0(\lambda ) & = & 0. &  &  & (b) 
\end{array}
\end{equation}
Extrema of the effective potential are given by roots of $I_{2-}=0$ with
scale-factors defined as 
\begin{equation}
\label{5.36}y^D=-\frac{2(D_1+1)\bar a_0(\lambda )}{\Lambda D_1}\equiv -\frac{%
2(D_1+1)Q_1(\lambda )}{\Lambda D_1}. 
\end{equation}
Substitution of (\ref{5.36}) into minimum-conditions (\ref{5.25c003}) yields
the following inequalities 
\begin{equation}
\label{5.37} 
\begin{array}{l}
Q_1(\lambda )\geq 0,\quad Q_2(\lambda )\geq 0,\quad Q_1(\lambda
)+Q_2(\lambda )>0 \\  
\\ 
8(D_1+1)(D_1+2)Q_1(\lambda )Q_2(\lambda )\geq (4D_1+5)^2(\lambda ^2-1)^2\bar
b_0^2(\lambda ). 
\end{array}
\end{equation}
From (\ref{5.36}) and (\ref{5.37}) we see that for even $D$ positive $y$ are
only allowed when the bare cosmological constant $\Lambda $ is negative: $%
\Lambda <0$.

As in the case of nonvanishing curvature scalars so also roots of $I_{2-}=0$
split into two classes. For nondegenerated physical relevant configurations $%
(\lambda \neq 1)$ the corresponding $\lambda _i$ must satisfy equation 
\begin{equation}
\label{5.38}\bar b_0(\lambda )=\sum_{i=0}^{D_1}P_i(2+i)\lambda
^{D_1-i}\sum_{j=0}^{i+1}\lambda ^{2j}=0. 
\end{equation}
For $\lambda >0$ this is only possible when there exist $P_i$ with different
signs.

In the case of degenerate configurations $(\lambda =1)$ equation $I_{2-}=0$
is trivially satisfied and the scale-factor at the minimum of the effective
potential given by 
\begin{equation}
\label{5.39}y_0^D=-\frac{4(D_1+1)\sum_{i=0}^{D_1}P_i}{\Lambda D_1}\ 
\end{equation}
with additional condition 
\begin{equation}
\label{5.40}\sum_{i=0}^{D_1}P_i\geq 0,\quad \sum_{i=0}^{D_1}P_i(2+i)^2\geq
0,\quad \sum_{i=0}^{D_1}P_i\left[ (2+i)^2+1\right] >0. 
\end{equation}
From inequalities (\ref{5.37}) and (\ref{5.40}) immediately follows that
effective potentials with parameters $(P_0<0,\ldots ,P_{D_1}<0)$ are not
stable.

{\bf - b,5) Vanishing curvature scalars and vanishing cosmological constants 
$(R_1=0,\ \Lambda =0)$:}

In this case equations (\ref{5.18}) contain only the projective coordinate $%
\lambda =y/x$%
\begin{equation}
\label{5.41} 
\begin{array}{llllllll}
I_{2+} & = & -2(D-2)\bar a_0(\lambda ) & = & 0 &  &  & (a) \\  
&  &  &  &  &  &  &  \\ 
I_{2-} & = & -2(\lambda ^2-1)\bar b_0(\lambda ) & = & 0. &  &  & (b) 
\end{array}
\end{equation}
Corresponding physical configurations are possible for domains in parameter
space given by 
\begin{equation}
\label{5.42}\lambda =1,\quad \sum_{i=0}^{D_1}P_i=0 
\end{equation}
or 
\begin{equation}
\label{5.43}\lambda \neq 1,\quad R_\lambda [\bar a_0(\lambda ),\bar
b_0(\lambda )]=0. 
\end{equation}
Minima of the effective potential are localized at lines $\{\lambda _i=y/x\}$
and must be stabilized by additional terms. Otherwise we get an unstable
''run-away'' minimum of the potential.

{\bf - c) Generalization to $n-$scale-factor models:}

The analytical methods used in the above considerations on stability
conditions of internal space configurations with two scale-factors can be
extended to configurations with 3 and more scale-factors by techniques of
the theory of commutative rings \cite{34}. In this case constraints, similar
to polynomial (\ref{5.11}) $w(\lambda )$, follow from resultant systems on
homogeneous polynomials. We note that in master equations (\ref{5.8})
%, (\ref{5.8a}) 
we can
pass from affine coordinates $\{x,\ y\}$ to projective coordinates $\{X,
Y, Z \mid  x=X/Z, y=Y/Z\}$ and transform \mbox{polynomials $I_{1\pm }$} to
homogeneous polynomials in $\{X,\ Y,\ Z\}$ so that these generalizations are
immediately to perform. A deeper insight in extremum conditions can be
gained by means of algebraic geometry \cite{35}. Polynomials $I_{1\pm }$
define two algebraic curves on the $\{x,\ y\}-$plane and solutions of system
(\ref{5.8})
%, (\ref{5.8a}) 
$I_{1\pm }(x,y)=0$ correspond to intersection-points of these
curves. For $n$ scale-factors extremum conditions $\{\partial
_{a_i}U_{eff}=0\}_{i=1}^n$ would result in $n$ polynomials $I_n(x_1,\ldots
,x_n)=0$ defining $n$ algebraic varieties on $\RR ^n$. The sets of
solutions of system $I_n(x_1,\ldots ,x_n)=0$, or equivalently, the
intersection points of the corresponding algebraic varieties, define the
extremum points of $U_{eff}$.

\section{Perfect fluid potentials\label{mark4}}

\setcounter{equation}{0}

In the case of a multicomponent perfect fluid the energy density reads \cite
{26,27} 
\begin{equation}
\label{6.1}\rho =\sum_{a=1}^m\rho ^{(a)}=\sum_{a=1}^mA_a\exp {\left(
-\sum_{i=1}^n\alpha _i^{(a)}D_i\beta ^i\right) }, 
\end{equation}
where $A_a$ are arbitrary positive constants. This formula describes an $m$
- component perfect fluid with equations of state $P_i^{(a)}=\left( \alpha
_i^{(a)}-1\right) \rho ^a$ in the internal space $M_i\ (i=1,\ldots ,n)$.
In the external space each component corresponds 
to vacuum: $\alpha _0^{(a)}=0 \ (a=1,\ldots ,m)$. 
Physical values of $\alpha _i^{(a)}$ are restricted to 
\begin{eqnarray}\label{6.2}
0\leq \alpha ^{(a)}_i\leq 2.
\end{eqnarray} 
It is easy to see that the case $\alpha _i^{(a)}=0 \ \forall a,i$
corresponds to the vacuum in spaces $M_i$ and contributes to the bare
cosmological constant $\Lambda $. 
Therefore we shall
not consider this case here, because it leads
to the pure geometrical potential of section \ref{mark2}. The other
limiting case \\ $\alpha _i^{(a)}=2\delta _i^{(a)},\ m=n$ formally coincides
with the ''monopole'' potential, which will be considered in the next
section. %We note that in a phenomenological way equations of state and
%energy density (\ref{6.1}) can represent free scalar fields living on the
%internal space components \cite{36b}.

{\bf -a) One-scale-factor model:}

Let us first analyze a one-component perfect fluid living in a
one-scale-factor model. In this case energy density (\ref{6.1}) reads $\rho
=Ae^{-\alpha D_1\beta }$ and for a vanishing effective cosmological constant 
$\Lambda _{eff}=0$ the extremum condition leads to 
\begin{equation}
\label{6.3}R_1e^{(\alpha D_1-2)\beta _c}=\kappa ^2\alpha D_1A 
\end{equation}
and 
\begin{equation}
\label{6.4}R_1e^{-2\beta _c}=\frac{2\alpha D_1}{\alpha D_1-2}\Lambda . 
\end{equation}
For the second derivative of the effective potential in the minimum we
obtain: 
\begin{equation}
\label{6.5}a_{11}=\left. \frac{\partial ^2U_{eff}}{\partial \beta ^2}\right|
_{\beta _c}=(\alpha D_1-2)R_1{\left( e^{-2\beta _c}\right) }^{\frac{D-2}{%
D_0-2}}. 
\end{equation}
Because of $\alpha ,A>0$, equation (\ref{6.3}) shows that the internal space 
$M_1$ should have positive curvature: $R_1>0$. From eq. (\ref{6.5}) we see
that there exists a minimum if $\alpha >2/D_1$. The corresponding mass
squared of the exciton is given as 
\begin{equation}
\label{6.6}m^2=\frac{(D_0-2)(\alpha D_1-2)}{D_1(D-2)}R_1{\left( e^{-2\beta
_c}\right) }^{\frac{D-2}{D_0-2}}. 
\end{equation}
For the critical value of $\alpha $ at $\alpha =2/D_1$ the model becomes
degenerated: $U_{eff}\equiv 0$.

As illustration, let $M_1$ be a 3 - dimensional sphere and $a_c=10L_{Pl}$.
This minimum can be achieved for $A={\left( \alpha \pi ^2\right) }^{-1}\cdot
10^{\alpha D_1-2}$. Thus, $\frac 3{2\pi ^2}<A\leq 5\cdot 10^2$ and $%
0<m^2\leq \frac{16}5\cdot 10^{-5}$ for $2/D_1<\alpha \leq 2$ and $D_0=4$.
So, conditions (\ref{3.19a}) are satisfied.

{\bf -b) Two-scale-factor model:}

For a multicomponent perfect fluid with energy density (\ref{6.1}) living in
a two-scale-factor model the effective potential reads 
\begin{equation}
\label{6.7} 
\begin{array}{ll}
U_{eff}= & {\left( \prod_{i=1}^2e^{D_i\beta ^i}\right) }^{-\frac
2{D_0-2}}\left[ -\frac 12\sum_{i=1}^2R_ie^{-2\beta ^i}+\Lambda +\right. \\  
&  \\  
& \left. +\kappa ^2\sum_{a=1}^mA_a\exp {\left( -\sum_{k=1}^2\alpha
_k^{(a)}D_k\beta ^k\right) }\right] . 
\end{array}
\end{equation}
Introducing the abbreviations 
\begin{equation}
\label{6.8} 
\begin{array}{l}
u_k^{(a)}:=\alpha _k^{(a)}+ 
\frac{2-\sum_{i=1}^2\alpha _i^{(a)}D_i}{D-2},\quad v_k^{(a)}:=\widetilde{h}%
_a\alpha _k^{(a)},\quad c_k:=\frac{2\Lambda D_k}{D-2}, \\  \\ 
h_a:=\kappa ^2A_ae^{-\alpha _1^{(a)}D_1 
{\beta _c^1}}e^{-\alpha _2^{(a)}D_2{\beta _c^2}}>0,{\ } \\  \\ 
\widetilde{h}_a:=h_a\exp {\left[ -\frac 2{D_0-2}\sum_{i=1}^2D_i\beta
_c^i\right] } 
\end{array}
\end{equation}
extremum condition and Hessian can be calculated to yield 
\begin{equation}
\label{6.9} 
\begin{array}{l}
\frac{\partial U_{eff}}{\partial \beta ^k}=0,\ k=1,2 \  
\Rightarrow \\  
\\ 
I_k:=c_k+D_k\kappa ^2\sum_{a=1}^mA_au_k^{(a)}e^{-\alpha _1^{(a)}D_1{\beta
_c^1}}e^{-\alpha _2^{(a)}D_2{\beta _c^2}}-R_ke^{-2\beta _c^k}=0 
\end{array}
\end{equation}
and 
\begin{equation}
\label{6.10} 
\begin{array}{ll}
a_{(c)ik} & \equiv \left. 
\frac{\partial ^2U_{eff}}{\partial \beta ^i\,\partial \beta ^k}\right|
_{\vec \beta _c} \\  &  \\  
& =-\frac{4\Lambda _{eff}}{D_0-2}\left[ \frac{D_iD_k}{D_0-2}+\delta
_{ik}D_k\right] +\sum_{a=1}^m\widetilde{h}_a{\alpha _k^{(a)}D_k\left( {%
\alpha _i^{(a)}D_i-2\delta _{ik}}\right) .} 
\end{array}
\end{equation}
From the auxiliary matrix 
\begin{equation}
\label{6.11}\left[ \bar G^{-1}A_c\right] _{ik}=-\frac{4\Lambda _{eff}}{D_0-2}%
\delta _{ik}+J_{ik},\quad
J_{ik}=\sum_{a=1}^mv_k^{(a)}(D_ku_i^{(a)}-2\delta _{ik}) 
\end{equation}
we get then the exciton masses squared as 
\begin{equation}
\label{6.12}m_{1,2}^2=-\frac{4\Lambda _{eff}}{D_0-2}+\frac 12\left[ Tr(J)\pm 
\sqrt{Tr^2(J)-4\det (J)}\right] . 
\end{equation}
Similar to the case of the Casimir-like potential, considered in 
\mbox{section \ref{mark3},} 
extremum condition (\ref{6.1}) has the form of a system of
equations in variables $z_1=e^{-{\beta _c^1}},\ z_2=e^{-{\beta _c^2}}$%
\begin{equation}
\label{6.13}I_k=c_k+D_k\kappa ^2\sum_{a=1}^mA_au_k^{(a)}z_1^{\alpha
_1^{(a)}D_1}z_2^{\alpha _2^{(a)}D_2}-R_kz_k^2=0,\quad k=1,2 
\end{equation}
and for a given point $p=\left\{ \Lambda ,R_1,R_2,\ A_1,\ldots ,A_m,\alpha
_1^{(1)},\ldots ,\alpha _2^{(m)}\right\} $ in parameter space $\RR
_{par}^{3(m+1)}$ positions of extrema should be found as solutions of this
system. In contrast with (\ref{5.8}), 
%(\ref{5.8a}), 
the powers of $z_i$ are real $(\alpha
_i^{(a)}\in \left[ 0,2\right] \subset \RR )$ so that in the general case the
solutions should be found numerically. Partially analytical methods can be
applied, e.g. for $\alpha _i^{(a)}\! $ 
rational ($\alpha _i^{(a)}\! \in \! \QQ $). In
this case the \mbox{representation 
$\alpha _i^{(a)}D_i=\frac{n_i^{(a)}}{d_i^{(a)}}$}
holds with natural numerator $n_i^{(a)}\in \NN $ and denominator $%
d_i^{(a)}\in \NN ^{+}$, and $n_i^{(a)},d_i^{(a)}$ relative prime, GCD$%
(n_i^{(a)},d_i^{(a)})=1$. Introducing the least common multiple of the
denominators $l=$LCM$(d_1^{(1)},...,d_2^{(m)})$ and the natural numbers $%
\vartheta _i^{(a)}:=\frac l{d_i^{(a)}}n_i^{(a)}$ one has $\alpha
_i^{(a)}D_i= \frac{\vartheta _i^{(a)}}l$. Eqs. (\ref{6.13}) transform then
to a system of polynomials 
\begin{equation}
\label{6.14}I_k=c_k+D_k\kappa ^2\sum_{a=1}^mA_au_k^{(a)}y_1^{\vartheta
_1^{(a)}}y_2^{\vartheta _2^{(a)}}-R_ky_k^{2l}=0,\quad k=1,2 
\end{equation}
in the new variables $y_k=z_k^{1/l}$, which can be analyzed by algebraic
methods \cite{34,35} and for rational parameters by methods of number theory 
\cite{36}. So, for common roots of equations $I_1=0,\ I_2=0$ the resultants 
\cite{34} $R_{y_1}\left[ I_1,I_2\right] ,\ R_{y_2}\left[ I_1,I_2\right] $
must necessarily vanish 
\begin{equation}
\label{6.15}R_{y_1}\left[ I_1,I_2\right] =w(y_2)=0,\ R_{y_2}\left[
I_1,I_2\right] =w(y_1)=0 
\end{equation}
and the analysis of (\ref{6.13}) can be reduced to an analysis of the
polynomials $w(y_1), w(y_2)$ of degree 
\begin{equation}
\label{6.16}\deg \left[ w(y_1)\right] ,\deg \left[ w(y_2)\right] \leq
\left[ l\stackunder{a}{\max }(\alpha _1^{(a)}D_1+\alpha
_2^{(a)}D_2,2)\right] ^2 
\end{equation}
in only one of the variables $y_1$ and $y_2$ respectively. In contrast with
equation system (\ref{5.8})
%, (\ref{5.8a}) 
for the Casimir-like potential, for arbitrary $%
\vartheta _i^{(a)}$ the sum-term in (\ref{6.14}) cannot be factorized and
the explicit calculation of resultants (\ref{6.15}) cannot be circumvented.
So, the further analysis should be performed by computer-algebraic programs.

We turn now to some concrete subclasses of perfect fluids, which allow
analytical considerations.

{\bf - b,1) $m-$component perfect fluid with $\alpha _i^{(a)}=\alpha ^{(a)}$:%
}

In this case there exist no massive excitons for vanishing effective
cosmological constants $\Lambda _{eff}=0$. Indeed, $m_{1,2}^2>0$ and eq. (%
\ref{6.12}) imply $Tr(J)>0,\ \det (J)>0$ which with 
\begin{equation}
\label{6.17}J_{ik}=D_kW_1-2\delta _{ik}W_2,\quad
W_1:=\sum_{a=1}^mu^{(a)}v^{(a)},\ W_2:=\sum_{a=1}^mv^{(a)} 
\end{equation}
read $Tr(J)=D^{^{\prime }}W_1-4W_2>0$, $\det (J)=2W_2(2W_2-D^{^{\prime
}}W_1)>0$. But because of $v^{(a)}=\widetilde{h}_a\alpha ^{(a)}>0$ and
hence $W_2>0$ this leads to a contradiction. Thus, for the existence of
massive excitons $m_{1,2}^2>0$ the effective cosmological constant must be
negative $\Lambda _{eff}<0$.

{\bf - b,2) One-component perfect fluid with $\alpha _1\neq \alpha _2$:}

Again massive excitons are possible for negative effective cosmological
constants $\Lambda _{eff}<0$ only. For $\Lambda _{eff}=0$ we have here at
one hand $\det (J)=-2v_1v_2\delta \frac{D_0-2}{D-2}>0,\ \delta :=D_1\alpha
_1+D_2\alpha _2-2$ and hence $\delta <0$. On the other hand from $Tr(J)>0$
follows $\delta (\alpha _1+\alpha _2-\frac{\delta +2}{D-2})>0$ and hence $%
0>(D_0-2)(\alpha _1+\alpha _2)+D_1\alpha _1+D_2\alpha _2$. Because of $%
\alpha _k>0$ this is impossible and so should be $\Lambda _{eff}<0$.

{\bf - b,3) One-component perfect fluid with $\alpha _1=\alpha _2=\alpha $:}

For this subclass of b,1) extremum conditions (\ref{6.9}) can be
considerably simplified to yield 
\begin{equation}
\label{6.18}h=\kappa ^2Ae^{-\alpha (D_1{\beta _c^1+}D_2{\beta _c^2)}}=\frac
1{(D_0-2)\alpha +2}\left( \frac{D-2}{D_k}R_ke^{-2\beta _c^k}-2\Lambda
\right) 
\end{equation}
and the same fine-tuning condition as in the case of a pure geometrical
potential 
\begin{equation}
\label{6.19}\tilde C=\frac{R_1}{D_1}e^{-2\beta _c^1}=\frac{R_2}{D_2}%
e^{-2\beta _c^2}. 
\end{equation}
An explicit estimation of exciton masses and effective cosmological constant
can be easily done. Using (\ref{6.8}), (\ref{6.12}), (\ref{6.17}) we rewrite
the exciton masses squared as 
\begin{equation}
\label{6.20} 
\begin{array}{ll}
\left( 
\begin{array}{c}
m_1^2 \\ 
m_2^2 
\end{array}
\right) = & \frac 1{D-2}\left\{ -4\Lambda +h\left[ (D_0-2)\alpha +2\right]
\left[ \left( 
\begin{array}{c}
D^{^{\prime }}\alpha \\ 
0 
\end{array}
\right) -2\right] \right\} \times \\  
&  \\  
& \times \exp {\left[ -\frac 2{D_0-2}\sum_{i=1}^2D_i\beta _c^i\right] } 
\end{array}
\end{equation}
and transform with (\ref{6.18}) inequalities $m_{1,2}^2>0,\ h>0$ to the
following equivalent condition 
\begin{equation}
\label{6.21}\frac 2{D-2}\Lambda <\tilde C<0. 
\end{equation}
Hence stable space configurations with massive excitons are only possible
for internal spaces with negative curvature $R_k<0$. Reparametrizing $%
\Lambda $ according to (\ref{6.21}) as 
\begin{equation}
\label{6.22}\Lambda =\frac{D-2}2\left( \tilde C-\tau \right) , 
\end{equation}
with $\tau >0$ --- a new parameter, we get for exciton masses squared and
effective cosmological constant 
\begin{equation}
\label{6.23}\left( 
\begin{array}{c}
m_1^2 \\ 
m_2^2 
\end{array}
\right) =\left[ \left( 
\begin{array}{c}
D^{^{\prime }}\alpha \tau \\ 
0 
\end{array}
\right) -2\tilde C\right] \exp {\left[ -\frac 2{D_0-2}\sum_{i=1}^2D_i\beta
_c^i\right] }, 
\end{equation}
\begin{equation}
\label{6.24}\Lambda _{eff}=-\frac{D_0-2}2\left[ \tau \frac{(D-2)\alpha }{%
(D_0-2)\alpha +2}-\tilde C\right] \exp {\left[ -\frac
2{D_0-2}\sum_{i=1}^2D_i\beta _c^i\right] .} 
\end{equation}
According to definition (\ref{6.22}) and equations (\ref{6.18}), (\ref{6.19}%
) the parameter $\tau $ can be expressed in terms of $\tilde C$ and $R_k$ as 
\begin{equation}
\label{6.25}\tau =\kappa ^2A\frac{(D_0-2)\alpha +2}{D-2}\left| \tilde
C\right| ^{\frac{D^{^{\prime }}\alpha }2}\prod_{k=1}^2\left| \frac{D_k}{R_k}%
\right| ^{\frac{D_k\alpha }2}. 
\end{equation}
Comparison of equations (\ref{6.23}), (\ref{6.24}) with formula (\ref{4.4})
shows that for $\tau \ll \tau _0\equiv \left| \tilde C\right| \min (\frac
2{D^{^{\prime }}\alpha },\frac{(D_0-2)\alpha +2}{(D-2)\alpha })$ we return
to the pure geometrical potential considered in section \ref{mark2}. So
physical conditions (\ref{3.19a}) are fulfilled for internal space
configurations with sufficiently high dimensions greater then some critical
dimension $D_{crit}$. From (\ref{6.23}) and (\ref{6.24}) we see that
depending on the value of $\tau $ this critical dimension $D_{crit}$ can
only be larger then that for the pure geometrical model. According to (\ref
{6.25}) there exist excitons for any positive and finite values of the fluid
parameter $A$, but than larger $A$ for fixed $\alpha $ than larger would be
the critical dimension $D_{crit}$. (Here we take into account that $\kappa
^2=\mu $ and that the volume $\mu $ of the compact internal factor spaces
with constant negative curvature is finite.)

Comparing the results of this subsection with the results for the
one-scale-factor model at the beginning of the section we see that there
exists a different behavior of the perfect fluid models in the case of
vanishing effective cosmological constant $\Lambda _{eff}=0$. For the
one-scale-factor model massive excitons are allowed for $\Lambda _{eff}=0$,
whereas in the two-scale-factor model they cannot occur. This means that,
according to the explanations of section \ref{mark1}, the $\Lambda _{eff}=0$
extremum of the effective potential must be a saddle point and we are
explicitly confronted with the specifics of the scale factor reduction
described by eq. (\ref{3.19}).

\section{''Monopole'' potentials}

\setcounter{equation}{0}

The ''monopole'' ansatz \cite{c2} consists in the proposal that the
antisymmetric tensor field $F^{(i)}$ of rank $D_i$ is not equal to zero
only for components corresponding to the internal space $M_i$. The energy
density of these fields reads \cite{10,11} 
\begin{equation}
\label{7.1}\rho =\sum_{k=1}^n(f_k)^2e^{-2D_k\beta ^k}, 
\end{equation}
where $f_k$ are arbitrary constants (free parameters of the model). Energy
density (\ref{7.1}) formally coincides with the energy density (\ref{6.1})
of a multicomponent perfect fluid with parameters $\alpha _i^{(a)}=2\delta
_i^{(a)},\ m=n$ and $A_k=(f_k)^2$, so that the calculations parallel that
of the previous section.

Extremum condition (\ref{3.1}) leads in the case of vanishing effective
cosmological constant $\Lambda _{eff}=0$ to a fine-tuning of the scale
factors 
\begin{equation}
\label{7.2}\frac{R_k}{2\kappa ^2D_k(f_k)^2}=e^{-2\beta ^k(D_k-1)} 
\end{equation}
and 
\begin{equation}
\label{7.3}\Lambda =\frac 12\sum_{k=1}^nR_ke^{-2\beta ^k}\frac{D_k-1}{D_k}\
, 
\end{equation}
so that extrema are only possible iff $R_k>0,\ \Lambda >0$.

For a one-scale-factor model the exciton mass squared reads 
\begin{eqnarray}\label{7.4}
m^2 = \frac{2(D_0-2)(D_1-1)}{D_1(D-2)}R_1
{\left (e^{-2\beta _c}\right)}^{\frac{D-2}{D_0-2}}.
\end{eqnarray} Condition (i) is satisfied if 
\begin{eqnarray}\label{7.5}
f^2  \mbox{ \small $^{>}_{\sim}$ }\left.R_1\right/2\kappa
^2D_1.
\end{eqnarray}Let $M_1$ be a 3 - dimensional sphere, then $R_1=6$ and $%
\kappa ^2=2\pi ^2$. To get a minimum of the effective potential for a scale
factor $a_c=10L_{Pl}$ we should take $f^2\approx 5\cdot 10^2$. For this
value of $a_c$ and for $D_0=4$ the mass squared is $m^2=\frac{16}5\cdot
10^{-5}\ll M_{Pl}^2$. Thus, all three conditions (\ref{3.19a}) are satisfied.

For a two-scale-factor model the exciton masses are given 
\mbox{by (\ref{6.12})} 
\begin{equation}
\label{7.6}m_{1,2}^2=\frac 12\left[ Tr(J)\pm \sqrt{Tr^2(J)-4\det (J)}\right]
, 
\end{equation}
where in terms of abbreviations (\ref{6.8}) matrix $J$ reads 
\begin{equation}
\label{7.8}J_{ik}=4\widetilde{h}_k(D_k-1)\left[ \delta _{ik}-\frac{D_k}{D-2}%
\right] . 
\end{equation}

One immediately verifies that $Tr(J)>0,\ \det (J)>0, \\ Tr^2(J)-4\det
(J)\geq 0$ for dimensions $D_1>1,\ D_2>1$ and hence \\ $0<m_2^2\leq \frac
12Tr(J)\leq m_1^2<Tr(J)$. This means that physical conditions (\ref{3.19a})
are satisfied if $Tr(J)\leq M_{Pl}^2$ and $e^{\beta _c^k}$%
\mbox{ \small
$^{>}_{\sim}$ }\thinspace $L_{Pl}$. Substituting 
\begin{equation}
\label{7.9}\widetilde{h}_k=\frac{R_k}{2D_k}e^{-2\beta _c^k}\exp {\left[
-\frac 2{D_0-2}\sum_{i=1}^2D_i\beta _c^i\right] } 
\end{equation}
into (\ref{7.8}) we get the matrix trace as 
\begin{equation}
\label{7.10} 
\begin{array}{ll}
Tr(J)= & \frac 2{D-2}\left[ 
{\sum_{k=1}^2\frac{(D_k-1)}{D_k}R_k(D-2-D_k)}e^{-2\beta _c^k}\right] \times
\\  &  \\  
& \times \exp {\left[ -\frac 2{D_0-2}\sum_{i=1}^2D_i\beta _c^i\right] .} 
\end{array}
\end{equation}
With this formula at hand we have e.g. for an internal space configuration $%
M_1\times M_2:\ M_1=$ $S^3,\ a_{(c)1}=10L_{Pl}\ ;\ \ M_2=$ $S^5,\
a_{(c)2}=10^2L_{Pl}$ the\ estimate $Tr(J)\approx 56\cdot 10^{-14}M_{Pl}^2\ll
M_{Pl}^2$ and all conditions (i) - (iii) of (\ref{3.19a}) are satisfied.

\section{Cosmological stability of compactified internal space configurations
}

\setcounter{equation}{0}

In section \ref{mark1} we showed that inhomogeneous scale factor
fluctuations of internal factor spaces have the form of massive scalar
fields in the external space-time. These scalar fields are coupled with the
gravitational field $\hat g_{\mu \nu }^{(0)}$ of the external space-time.
Thus the energy of the external gravitational field can enlarge scalar field
perturbations during the universe evolution. The type of reasoning that was
used by Maeda \cite{12} shows that for an expanding external space scalar
field perturbations decrease with time.

To show it we consider an external space-time metric in the form 
\begin{equation}
\label{8.1}\hat g^{(0)}=-e^{2\hat \gamma (\tau )}d\tau \otimes d\tau +\hat
a_0^2(\tau )\bar g^{(0)},
\end{equation}
where $\hat a_0=e^{\hat \beta ^0}$ is the scale factor of the external space
in the Einstein frame and $\bar g^{(0)}$ is $D_0$-dimensional constant
curvature space : 
\begin{equation}
\label{8.2}R\left[ \bar g^{(0)}\right] =kD_0\left( D_0-1\right) \equiv R_0\
,k=\pm 1,0.
\end{equation}
Since we are interested in cosmological solutions, we restrict our model to
homogeneous scalar fields. The behavior of such model is described by the
Lagrangian
\begin{equation}
\label{8.3}
\begin{array}{ll}
L= & e^{\hat \gamma }e^{D_0\hat \beta ^0}\left\{ e^{-2\hat \beta
^0}R_0+e^{-2\hat \gamma }D_0\left( 1-D_0\right) \left( \dot{\hat \beta ^0}\right)
^2+\right.  \\  
&  \\  
& \left. +e^{-2\hat \gamma }\sum_{i=1}^n\left( \dot \varphi \right)
^2-2U_{eff}\right\} +2D_0\frac d{d\tau }\left( e^{-\hat \gamma }e^{D_0\hat
\beta ^0}\dot{\hat \beta ^0}\right) 
\end{array}
\end{equation}
with constraint
\begin{equation}
\label{8.4}
\begin{array}{ll}
\frac{\partial L}{\partial \hat \gamma }=0\Rightarrow \rho  & =\frac
12e^{-2\hat \gamma }\sum_{i=1}^n\left( \dot \varphi ^i\right) ^2+U_{eff} \\  
&  \\  
& =\frac 12\left[ e^{-2\hat \beta ^0}R_0+D_0\left( D_0-1\right) e^{-2\hat
\gamma }\left( \dot{\hat \beta ^0}\right) ^2\right] ,
\end{array}
\end{equation}
where the overdot denotes differentiation with respect to time $\tau $ \ and 
$\rho =-T_0^0$ is the energy density of the system. The equations of motion
for the scalar fields are 
\begin{equation}
\label{8.5}
\ddot \varphi ^i+(D_0\dot{\hat \beta ^0}-\dot{\hat \gamma })\dot \varphi
^i+e^{2\hat \gamma }\frac{\partial U_{eff}}{\partial \varphi ^i}=0.
\end{equation}
It can be easily seen that the energy density $\rho $ satisfies following
equation 
\begin{equation}
\label{8.6}
\dot \rho \equiv \frac{d\rho }{d\tau }=-D_0\dot{\hat \beta ^0}e^{-2\hat
\gamma }\sum_{i=1}^n\left( \dot \varphi ^i\right) ^2.
\end{equation}
We see that in a synchronous system $\hat \gamma =0$ of an expanding
external space 
$(H\equiv \dot{\hat \beta ^0}=\frac{\dot{\hat a_0}}{\hat a_0}>0)$, the energy
density $\rho $ decreases with time. Thus in the comoving system (Einstein
frame) our model is always dissipative and the universe can reach the
effective potential minimum.

Additionally to $\rho $ we can consider the quantity $E=v_0\rho $, where $%
v_0=e^{D_0\hat \beta ^0}$ is proportional to the volume of the external
space. For closed external spaces $E$ plays the role of a total energy,
which varies in time as
\begin{equation}
\label{8.7}
\begin{array}{ll}
\dot E & =D_0\dot{\hat \beta ^0}\left( E-v_0e^{-2\hat 
\gamma }\sum_{i=1}^n\left(
\dot \varphi ^i\right) ^2\right)  \\  
&  \\  
& =D_0\dot{\hat \beta ^0}v_0\left( -\frac 12e^{-2\hat 
\gamma }\sum_{i=1}^n\left(
\dot \varphi ^i\right) ^2+U_{eff}\right) ,
\end{array}
\end{equation}
and near minima according to (\ref{3.6}) as 
\begin{equation}
\label{8.8}
\dot E=D_0\dot{\hat \beta ^0}v_0\left( \Lambda _{eff}-\frac 12e^{-2\hat
\gamma }\sum_{i=1}^n\left( \dot \psi ^i\right) ^2+\frac
12\sum_{i=1}^nm_{(c)i}^2(\psi ^i)^2\right) .
\end{equation}
In contrast with (\ref{8.6}) the total energy can for $\dot{\hat \beta ^0}>0$
decrease as well as increase.

Relevant for directly observable physical characteristics like cross
sections, transition probabilities etc. is the energy density $\rho =-T_0^0$%
. It also enters the Einstein equations and defines by this way the dynamics
of the scale factor $\hat a_0=e^{\hat \beta ^0}$ of the external space. The
total energy $E$ is not less interesting, because it is connected with the
Wheeler-deWitt equation and plays a role in a quantized midisuperspace model.

\section{Conclusions}

\setcounter{equation}{0}

In the present lecture we studied stability conditions for compactified
internal spaces. Starting from a multidimensional cosmological model we
performed a dimensional reduction and obtained an effective four-dimensional
theory in Brans - Dicke and Einstein frames. The Einstein frame was
considered here as the physical one \cite{37}. In this frame we derived an
effective potential. It was shown that small excitations of the scale
factors of internal spaces near minima of the effective potential have a
form of massive scalar particles (gravitational excitons) developing in the
external space - time. The exciton masses strongly depend on the dimensions
and curvatures of the internal spaces, and possibly present additional
fields living on the internal spaces. These fields will contribute to the
effective potential, e.g. due to the Casimir effect, and by this way affect
the dynamics of the scale factor excitations. So, the detection of the scale
factor excitations can not only prove the existence of extra dimensions, but
also give additional information about the dimension of the internal spaces
and about fields possibly living on them.

For some particular classes of effective potentials with one, two and $n$
scale factors we calculated exciton masses as functions of parameters of the
internal spaces and derived stability criterions necessary for the
compactification of the spaces.

Our analysis shows that conditions for the existence of stable
configurations may depend not only on dimension and topology of the internal
spaces, and additional fields contributing to the effective potential, but
also on the number of independently oscillating scale factors. For example, $%
n-$scale-factor models with a saddle point as extremum of the effective
potential $U_{eff}(\beta _1,\ldots ,\beta _n)$ would lead to an unstable
configuration. Masses of the corresponding excitations would be positive
(excitons) as well as negative (tachyons). Under scale factor reduction to
an $m-$scale factor model with $m<n$, i.e. when we connect some of the scale
factors by constraints $\beta _i=\beta _k$, the saddle point may, for
certain potentials, reduce to a stable minimum point of the new effective
potential $U_{eff}(\beta _1,\ldots ,\beta _m)$. As result all masses of
excitations would be positive (excitons). We demonstrated this
''stabilization via scale factor reduction'' explicitly on a model with
one-component perfect fluid.

In the present lecture we did not consider the case of degenerated minima of
the effective potential, for example, self - interaction - type potentials
or Mexican - hat - type potentials. In the former case one obtains massless
fields with self - interaction. In the latter case one gets massive fields
together with massless ones. Here, massless particles can be understood as
analog of Goldstone bosons. This type of the potential was described in \cite
{19}.

Another possible generalization of our model consist in the proposal that
the additional potential $\rho $ may depend also on the scale factor of the
external space. It would allow, for example, to consider a perfect fluid
with arbitrary equation of state in the external space. \bigskip

%%%%%%%%%%%%%%%%%%%%%%%%%%%%%%%%%%%%%%%%%%%%%%%%%%%%%%%%%%%%%%%%%
\newpage
{\bf Acknowledgements}\\ AZ thanks E.Guendelman and A.Kaganovich for their
cordial reception during the conference and the participants for interesting
discussions. UG acknowledges financial support from DAAD (Germany).

\end{document}